\documentclass[aip,reprint,onecolumn,english,nofootinbib,letterpaper,nobalancelastpage]{revtex4}
\usepackage{lmodern}

\usepackage[T1]{fontenc}
\usepackage[latin9]{inputenc}
\usepackage[letterpaper]{geometry}
\usepackage{amsmath}
\usepackage{amssymb}
\usepackage{bbm}
\usepackage{babel}
\usepackage{graphicx}

\usepackage{latexsym}
\usepackage{mathptm}

\newcommand{\f}{\begin{equation}}
\newcommand{\ff}{\end{equation}}

\begin{document}
\title{On the fate of Lorentz symmetry in relative-locality momentum spaces}

\author{Giovanni Amelino-Camelia}
\affiliation{\footnotesize{Dipartimento di Fisica, Universit\`a ``La Sapienza", P.le~Moro~2,~Roma,~EU}}
\affiliation{\footnotesize{INFN, Sez.~Roma1, P.le Moro 2, 00185 Roma, EU}}

\begin{abstract}
The most studied doubly-special-relativity scenarios,
 theories with both the speed-of-light scale
and a length/inverse-momentum
scale as non-trivial relativistic invariants, have concerned the possibility of
enforcing relativistically some nonlinear
laws on momentum space. For the  ``relative-locality framework"  recently
proposed in arXiv:1101.0931 a central role is played by  nonlinear laws
 on momentum space, with the guiding principle that they should provide
a characterization of the geometry of momentum space. Building on previous
doubly-special-relativity results I here identify a criterion for establishing
whether or not a given geometry of the relative-locality momentum space
is ``DSR compatible", {\it i.e.} compatible
with an observer-independent formulation of theories
on that momentum space. I find that given some chosen parametrization of momentum-space geometry the
criterion takes the form of an elementary algorithm.
I show that relative-locality momentum spaces that fail my criterion definitely ``break" Lorentz
invariance, {\it i.e.} theories on such momentum spaces necessarily are observer-dependent ``ether"
theories. By working out a few examples I provide evidence
that when the criterion is instead satisfied one does manage to produce a relativistic formulation.
The examples I use to illustrate the applicability of my criterion also have
 some intrinsic interest, including
 two particularly noteworthy cases of $\kappa$-Poincar\'e-inspired
momentum spaces.
 \end{abstract}

\maketitle

\vskip -0.95cm

\tableofcontents

\section{Introduction}
Freidel, Kowalski-Glikman, Smolin and I recently proposed~\cite{prl,grf2nd}
a class of theories which are most primitively formulated on momentum space,
and whose main characteristics are codified in the geometry of momentum space.
In that framework the momentum-space metric would primarily govern 
the form of the on-shell (``dispersion") relation, whereas the momentum-space
affine connection governs the law of composition of momenta (and ultimately
the law of conservation of ``total" momentum in processes).
The class of theories one might consider from this perspective
is evidently very wide, especially since it appears legitimate~\cite{prl,grf2nd}
to consider choices of affine connection on momentum space which are not Levi-Civita connections
(allowing for non-metricity and/or torsion on momentum space).
Exploring these possibilities is motivated in part by the fact
 that several
arguments in the quantum-gravity literature
could be viewed~\cite{prl,grf2nd}  as pointing toward a role for momentum-space
geometry. This was already remarkably emphasized
in Born's 1938 proposal~\cite{born1938} of a role for momentum-space curvature in the
study of the quantum-gravity problem, and has more recently resurfaced
through several independent arguments (see, {\it e.g.},
Refs.~\cite{majidCURVATURE,girelliCURVATURE,schullerCURVATURE}).
Moreover, even setting aside the possible role in the study of the quantum-gravity problem,
it appears that this novel framework deserves some interest because it provides
an opportunity for studying systematically, including a description of interactions among particles,
the possibility of a ``relativity of spacetime locality",
which had been previously confined~\cite{bob,leeINERTIALlimit,arzkowaRelLoc}
(also see the more recent Refs.~\cite{kappabob,transverse})
to the narrow scopes of simple
theories of free particles.

In the investigation of this ``relative-locality framework" of Refs.~\cite{prl,grf2nd}
it appears likely that,
both conceptually~\cite{prl,grf2nd} and from the viewpoint of phenomenology~\cite{leelaurentGRB,anatomy},
an important role will be played by the understanding of the fate of Lorentz
symmetry in theories formulated on the relative-locality momentum spaces.
And this is the main focus of the study I am here reporting.
I take as starting point the bulk of results on ``doubly-special-relativity (DSR) theories"
(see, {\it e.g.},
Refs.~\cite{dsr1Edsr2,jurekdsr1,dsrPOLAND2001,leeDSRprd,jurekDSR2,leeDSRrainbow,gacdsrrev2010}).
These are relativistic theories
 with both the speed-of-light scale
and a length/inverse-momentum
scale as non-trivial relativistic invariants,
and their most studied
formulations~\cite{dsr1Edsr2,jurekdsr1,dsrPOLAND2001,leeDSRprd,jurekDSR2,leeDSRrainbow,gacdsrrev2010}
concern indeed achieving a (deformed-)relativistic formulation
of physical laws introducing novel nonlinearities in momentum space.
In particular, the most studied such DSR scenarios allow for novel nonlinearities,
governed by the additional (length/inverse-momentum) relativistically-invariant scale,
in the on-shell relation and in the law of conservation of ``total" momentum,
so they provide a very close starting point for the investigation of the fate
of Lorentz symmetry in relative-locality momentum spaces, where the conjectured
non-trivial geometry of momentum space indeed primarily results~\cite{prl,grf2nd}
in modifications
of the on-shell relation and of the law of momentum conservation.

The most significant result I am here reporting is contained in Section~\ref{goldenrulesec}.
I introduce two requirements that must be satisfied by the metric and the affine connection
on momentum space in order for that momentum-space to be possibly ``DSR compatible",
{\it i.e.} such that the introduction of a characteristic scale of the geometry of momentum space
still allows the formulation of
observer-independent laws of physics.
I observe that for any given parametrization of momentum-space geometry my two requirements
can be expressed in terms of a simple algorithm, which in turn proves useful
for assessing very easily whether a given choice of momentum-space geometry
(a given choice of the parameters) can be DSR-compatible.
I show that when the geometry of a relative-locality momentum
space does not satisfy my two requirements then necessarily theories formulated
on that momentum space will require the introduction of a preferred ``ether" frame, a preferred class
of inertial observers.
I am unable to offer definite assurance that in all cases when instead the requirements
are satisfied it will be possible to achieve a DSR-compatible formulation of theories
on that momentum space: the requirements are evidently necessary but not so evidently sufficient.
Still by working out (in Secs.~\ref{dsr1sec} and \ref{torsionsec}) 
a few specific illustrative examples I provide some evidence
that my requirements might also be sufficient: the illustrative examples I consider
are such that the requirements are satisfied and for them I do manage to work out explicitly
a formulation of relativistic kinematics which does not require a preferred frame.

Before this main part of the manuscript contained 
in Secs.~\ref{goldenrulesec}, \ref{dsr1sec} and \ref{torsionsec},
I offer in Sec.~\ref{dsrgeneral} a brief review of the general structure of the
connection that DSR compatibility establishes between the form of the on-shell relation
and the form of the law of energy-momentum conservation.\\
Sec.~\ref{newmetric} contains a few remarks on how these results might affect the development
of the relative-locality framework.\\
Sec.~\ref{theoremsec} looks back at the criteria for DSR compatibility introduced
in Sec.~\ref{goldenrulesec} and frames them within a simple relativistic theorem.\\
The closing Sec.~\ref{closingsec} summarizes the main
results here obtained and attempts to identify some priorities for the next
steps that could be taken in this research area.

I denote the momentum-space relative-locality scale with $\ell$
(an inverse-momentum scale) and I work at leading order in $\ell$.
I am assuming that $|\ell|^{-1}$ is roughly of the order
of the huge Planck scale, so that a leading-order analysis might be all we
need for comparison to data we could realistically imagine to gather
over the next few decades.

\section{On-shellness, momentum conservation and deformations of Lorentz symmetry}\label{dsrgeneral}
Before proceeding with the main part of the analysis, let me pause briefly, in this section,
for summarizing the mains points originally made in Ref.~\cite{dsr1Edsr2}
concerning the consistency requirements
that the relativity of inertial frames imposes on the relationship between
the form of the on-shell(/dispersion) relation and the form 
of the law of energy-momentum conservation.\\
This is one of the most used DSR (``deformed special" , or ``doubly special", relativity)
results, and plays a pivotal role in the analysis I report in the following sections.

I start this section by revisiting the
transition from Galilean Relativity to Special Relativity
since this can be of guidance for following then the logical structure of
the transition from  Special Relativity to DSR.

\subsection{Aside on the transition from Galilean Relativity to Special relativity}
Galilean Relativity enforced the relativity of rest (and the associated relativity
of velocities). This is a notion that can be formulated without using a
reference scale, so in turn the transformation laws from a Galilean-inertial observer to another
do not involve any such reference scale.\\
This is manifest in all laws, and in particular in the on-shell relation
$$E= \frac{{\bf p}^2}{2m} (+m)~,$$
in the law of composition of velocities
$${\bf u} \oplus {\bf v} = {\bf u} + {\bf v}~,$$
which we use in particular when connecting the description of a velocity  ${\bf v}$
for a given observer Alice to the one of an observer Bob, when the relative Alice-Bob velocity
is ${\bf u}$, and in the law of composition of energy-momentum
$$p_\mu  \oplus p^\prime_\mu = p_\mu + p^\prime_\mu ~,$$
which we use in particular  to enforce energy-momentum conservation
when processes involving momentum exchange occur.

The transition from Galilean Relativity to Special Relativity
enforces the relativity of simultaneity and the associated law of absoluteness
of the speed of light. This does not challenge in any way the
Galilean law of energy-momentum composition,
which is indeed maintained $p_\mu  \oplus p^\prime_\mu = p_\mu + p^\prime_\mu$. But the role
of the speed of light as a relativistic invariant impose a change
of on-shell relation
$$E= \sqrt{c^2{\bf p}^2+c^4 m^2}~,$$
and a change in the law of composition of velocities
\begin{equation}
{\bf u} \oplus {\bf v} = \frac{1}{1+\frac{{\bf u} \cdot {\bf v}}{c^2}}
\left({\bf u} + \frac{1}{\gamma_u}  {\bf v}
+ \frac{1}{c^2} \frac{\gamma_u}{1+\gamma_u} ({\bf u} \cdot {\bf v}){\bf u}  \right)~,
\label{ungarVEL}
\end{equation}
where as usual $\gamma_u \equiv 1/\sqrt{1- {\bf u} \cdot {\bf u}/c^2}$.\\
Textbooks for undergraduates often choose to spare students the complexity
of the composition law (\ref{ungarVEL}), limiting the discussion to the
special case of (\ref{ungarVEL}) which occurs when ${\bf u}$ and ${\bf v}$ are collinear:
$${\bf u} \oplus {\bf v} \Big|_{collinear}
= \frac{{\bf u} + {\bf v}}{1+\frac{{\bf u} \cdot {\bf v}}{c^2}}~.$$
But the complexity of (\ref{ungarVEL}),
which is non-commutative and non-associative,
is well understood~\cite{ungar,ungarFOLLOWER,florianeteraUNGAR}
as playing a central role in the logical consistency of Special Relativity.
The special-relativistic prescription (\ref{ungarVEL}) renders the law of composition
of velocities compatible with the principle of relativistic invariance of the speed of light $c$.
Evidently insisting on the Galilean law ${\bf u} \oplus {\bf v} = {\bf u} + {\bf v}$ would have
been inconsistent with the presence of an invariant velocity scale: boost transformations of velocity
such that they saturate at $c$ could not possibly
admit ${\bf u} \oplus {\bf v} = {\bf u} + {\bf v}$ as an observer-independent prescription.
And of course the composition law (\ref{ungarVEL}) encodes all the richness of
special-relativistic boosts, including the Thomas-Wigner rotations (essentially the fact
that in the Lorentz algebra the commutator of non-parallel boosts produces a spatial rotation).

\subsection{DSR-compatibility between on-shellness and momentum conservation}
The idea of DSR-deformed Lorentz transformations was put forward~\cite{dsr1Edsr2} as a possible
description of certain {\underline{preliminary}}
theory results suggesting that there {\underline{might}} be violations
of some special-relativistic laws in certain approaches to the quantum-gravity problem,
most notably the ones based on spacetime noncommutativity and loop quantum gravity.
The part of the quantum-gravity community interested in those results was interpreting them
as a manifestation of a full breakdown of Lorentz symmetry, with the emergence of
a preferred class of observers (an ``ether"). But it was argued in ~\cite{dsr1Edsr2}
that departures from Special Relativity governed by a high-energy/short-distance scale
may well be compatible with the Relativity Principle, the principle of relativity
of inertial observers, at the cost of allowing some consistent modifications
of the Poincar\'e transformations, and particularly of the Lorentz-boost transformations.

The main area of investigation of the DSR proposal has been for the last decade
the possibility of introducing relativistically some deformed on-shell relations.
The  DSR proposal was put forward~\cite{dsr1Edsr2} as a conceptual path for pursuing
a broader class of scenarios of interest for fundamental physics, and in particular for
quantum-gravity research, including the possibility of introducing the second
observer-independent scale primitively in spacetime structure or primitively at the
level of the (deformed) de Broglie relation between wavelength and momentum.
However, the bulk of the preliminary results from quantum-gravity research
concern departures from the special-relativistic on-shell relation, and this in turn became
the main focus of DSR research.

My objective in this section is to remind my readers about the line of analysis,
originally discussed in Ref.~\cite{dsr1Edsr2}, which allows us to conclude
that {\underline{if}} the on-shell relation involves a relativistically-invariant
energy scale $M_* \equiv \ell^{-1}$ {\underline{and if}}
the relativity of inertial frames is to be preserved,
{\underline{then}} the scale $\ell$ must also intervene to modify the
law of composition of momenta. For my purposes here it indeed suffices to work in leading order
in $\ell$ (the scale $c$ from now on is set to $1$), focusing for simplicity
on the case of a 1+1-dimensional momentum space, and considering for definiteness
only {\underline{the illustrative example of}}  {\underline{a specific $\ell$-deformed
on-shell relation}}.
So let me consider the specific example of the on-shell relation\footnote{If $\ell$ is
of the order of the inverse of the Planck scale, the effects of DSR-deformed
Lorentz symmetry
in cases where the leading order is $\ell^2$ suppressed appear to be too soft to be
appreciated experimentally, even setting a time scale of a few decades from now.
This is the main reason why in the remainder of this manuscript I focus on cases
where the leading order is only suppressed linearly by $\ell$.
In this section however I am merely setting the stage for what follows by establishing
the logical connection between the on-shell relation and the
law of composition of momenta
in a relativistic theory, so it does no arm to contemplate a deformation
as weak as the one shown in Eq.~(\ref{dsr1quad}).}
\begin{equation}
m^2 = p_0^2 - p_1^2 - \ell^2 p_0^2 p_1^2 ~.
\label{dsr1quad}
\end{equation}
Evidently this law is not Lorentz invariant. If we insist on this law and on
the validity of classical (undeformed) Lorentz transformations between inertial
observers we clearly end up with a preferred-frame picture, and the Principle
of Relativity of inertial frames must be abandoned: the scale $\ell$ cannot
be observer independent, and actually the whole form of (\ref{dsr1quad}) is subject
to vary from one class of inertial observers to another.\\
The other option~\cite{dsr1Edsr2} in such cases is the DSR option of enforcing
the relativistic invariance of (\ref{dsr1quad}), preserving the relativity
of inertial frames, at the cost of modifying the action of boosts on momenta.
Then in such theories both the velocity scale $c$ (here mute only because of the
choice of dimensions) and the inverse-energy scale $\ell$ play the
same role~\cite{dsr1Edsr2}
of invariant scales of the relativistic theory which govern the form of boost
transformations. \\
Evidently {\underline{if}} the action of boosts on momenta is non-linearly deformed so that
 (\ref{dsr1quad})
 is invariant {\underline{one must then renounce to the}}  {\underline{linear
 law of composition
of momenta}}. In order to exhibit
some  formulas that illustrate this obvious fact, let me
introduce the following deformed boost action
\begin{equation}
[N, p_0] =  p_1 + \frac{3}{2} \ell^2 p_0^2 p_1 + \ell^2 p_1^3~,~~~
[N, p_1] =  p_0 + \frac{\ell^2}{2} p_0^3
\label{boostsquad}
\end{equation}
which evidently is such to leave invariant the deformed on-shell relation (\ref{dsr1quad}):
$$[N, p_0^2 - p_1^2 - \ell^2 p_0^2 p_1^2] =  0 ~.$$
Equally evident is the fact that these deformed boosts are relativistically
incompatible with the standard linear law of composition of momenta. Let us consider
for example the case of  a process with two incoming and two outgoing particles $a +b \rightarrow c+d$.
For this case one easily finds that
$$[N, (p^{(a)}+ p^{(b)})_\mu - (p^{(c)}+ p^{(d)})_\mu] \neq 0~,$$
even when $(p^{(a)}+ p^{(b)})_\mu = (p^{(c)}+ p^{(d)})_\mu$ is enforced.\\
Following the lessons of what turned out to be necessary for
the composition of velocities in going form Galilean Relativity to Special Relativity,
we can still look for laws of composition of momenta, $p \oplus p^\prime$,
that would be relativistically compatible with the deformed boosts.
A particular  example (actually a particularly simple example,
see later parts of this manuscript for other strategies of construction of the
composition law) is the following
\begin{eqnarray}
(p \oplus p^\prime)_0 &=&
p_0 + p^\prime_0 + \ell^2 p^{\prime}_0  p^{2}_1 +\ell^2 p_0  p^{\prime 2}_1
+ 2 \ell^2 p_1  p^{\prime}_1 (p_0 + p^{\prime}_0)
 \label{dsr1econsquad}\\
(p \oplus p^\prime)_1 &=&
p_1 + p^\prime_1 + \frac{\ell^2}{2}  p_1  p^{\prime 2}_0
+ \frac{\ell^2}{2}  p^{\prime}_1 p^{2}_0
+ \ell^2 p_0  p^{\prime}_0 (p_1 + p^{\prime}_1) ~.
\label{dsr1pconsquad}
\end{eqnarray}
With this prescription for the composition law the relativistic invariance
is restored; indeed one can easily verify that
when $ (p^{(a)} \oplus p^{(b)})_\mu= (p^{(c)} \oplus p^{(d)})_\mu$
one does have that
$$[N, (p^{(a)} \oplus p^{(b)})_\mu - (p^{(c)} \oplus p^{(d)})_\mu ] = 0~$$
(working again in leading order in $\ell^2$, consistently with the
approximations made above).

This completes my brief summary of the relativistic consistency
between modified on-shell relation and modified energy-momentum composition law
first observed in Ref.~\cite{dsr1Edsr2}.\\
This brief summary will suffice to prepare the intuition of the reader
for the results reported in the following.\\
I shall not dwell here on other aspects of DSR research,
which however I want to briefly bring to the attention of the reader before closing this section.\\
It was recently realized that in at least some DSR frameworks
the counterpart of the acquired absoluteness of the energy scale $\ell^{-1}$
is an acquired relativity of spacetime
locality~\cite{bob,leeINERTIALlimit,arzkowaRelLoc,kappabob,transverse}.
This fits naturally with the observation that in Special Relativity
the counterpart of the acquired absoluteness of the velocity scale $c$
is an acquired relativity of simultaneity (colloquially ``relative time").
From the DSR perspective the ``relative-locality framework" adopted in the
next sections is a particularly promising candidate for organizing logically,
in terms of the geometry of momentum space,
this relativity-locality features preliminarily characterized
in Refs.~\cite{bob,leeINERTIALlimit,arzkowaRelLoc,kappabob,transverse},
but of course the relative-locality framework may be considered even without
DSR-compatibility, as a powerful formalism that can be applied also to cases
where a preferred frame does arise.\\
Another framework that has been much considered from the DSR perspective if
the one of Hopf-algebra symmetries (see, {\it e.g.}, Refs.~\cite{lukieIW,majidruegg,kpoinap})
 which in particular naturally accommodates nonlinear actions of the
 type here shown in Eq.~(\ref{boostsquad}).\\
Another area of active research concerns the DSR description
of the properties of macroscopic bodies, composites
of a large number of microscopic particles, as discussed in particular in Ref.~\cite{dsrPOLAND2001}
(also see Ref.~\cite{soccerball} for the relative-locality-framework perspective).

\section{A ``relativistic golden rule" for the geometry
of momentum space}\label{goldenrulesec}
In this section I propose two criteria which can be used to establish
whether or not a given relative-locality momentum space is ``DSR compatible",
{\it i.e.} whether or not it is possible to formulate
relativistic theories (observer-independent laws) on
that momentum space.
The two criteria are labeled,
for reasons that shall soon be evident, ``no-photon-decay-switch-on constraint"
and ``no-pair-production-switch-off constraint".
It will be clear for the careful reader that these criteria
are applicable to any parametrization of the geometry of momentum
space, but for definiteness and simplicity my explicit derivations
are focused on a parametrization of the geometry of momentum space
which is not completely general.
For the momentum space metric I assume that it is such that
(in the sense of Refs.~\cite{prl,grf2nd})
the on-shell (``dispersion") relation takes the form
\begin{equation}
m^2 = p_0^2 - p_j^2 + \alpha_1 p_0 p_j^2 + \alpha_2 p_0^3 ~.
\label{metricMASTER}
\end{equation}
While for momentum-space affine connection
I assume that it is such that
(again in the sense of Refs.~\cite{prl,grf2nd})
the deformed law of composition of momenta
 takes the form
\begin{eqnarray}
(k \oplus p)_0 &=& k_0 + p_0 + \beta_1 {\vec k} \cdot {\vec p}
 + \beta_2 k_0 p_0
~, \label{connectionMASTER0}\\
(k \oplus p)_j &=& k_j + p_j + \gamma_1 k_0 p_j + \gamma_2 p_0 k_j ~.
\label{connectionMASTERj}
\end{eqnarray}
This is a 6-parameter family of (leading-order) momentum-space geometries,
with $\alpha_1,\alpha_2,\beta_1,\beta_2,\gamma_1,\gamma_2$
all assumed to be either of order $\ell$ or completely negligible with
respect to $\ell$.

The only other (however very weak) assumption needed for
the derivation of my ``golden rule"
is that the relevant momentum-space
theories involve a ``vertex/interaction term"~\cite{prl}
potentially eligible for photon decay $\gamma \rightarrow e^+ e^-$
\begin{equation}
 p_\gamma = p_+ \oplus p_-
\label{vertexgamma}
\end{equation}
and a ``vertex term" eligible for electron-positron pair production
 from photon-photon collisions $\gamma \, \Gamma \rightarrow e^+ e^-$
\begin{equation}
 p_{{\!\!~}_\Gamma} \oplus
 p_\gamma = p_+ \oplus p_-
\label{vertexgammagamma}
\end{equation}

\subsection{No-photon-decay-switch-on constraint}
The first ingredient of my proposed `golden rule"
is a ``no-photon-decay-switch-on constraint", essentially amounting to
the request that massless particles should not decay\footnote{While the second half of my criterion,
the requirement discussed in the next subsection (II.B) is completely new, the
requirement discussed in this subsection II.A, which is the first half
of my criterion, has been used in some previous studies of theories with
nonlinearities in momentum space, starting with the analyses reported
in Refs.~\cite{gacnewjourn,sethmajor}. Those previous studies however only used
this criterion for addressing specific issues within a given proposal of nonlinearities
in momentum space, whereas here I am using this requirement as a way to investigate
the fate of Lorentz symmetry on a large class of theories.
Another point to be stressed concerns the emphasis on photon decay:
focusing on photon decay is sufficient for my purposes, and allows
me keeps the discussion very explicit, but everything I observe
here for photon decay applies equally well to
some other ``forbidden decays" (processes forbidden in special relativity, which might
become allowed if Lorentz symmetry is broken) of ``light", but not necessarily massless, particles,
such as neutrinos.
An example of recent interest~\cite{whataboutopera} involving neutrinos,
specifically the process $\nu_\mu \rightarrow \nu_\mu e^+ e^-$,
was discussed in Ref.~\cite{operaDSR}.}.
The photon-decay
process $\gamma \rightarrow e^+ e^-$ is forbidden
when the geometry of momentum space is Minkowski/flat.
Relative-locality momentum spaces must be such that~\cite{prl,grf2nd}
for low-energy particles the geometry of momentum
space is indistinguishable from the Minkowski/flat case,
so in any relative-locality momentum space
low-energy photons will be stable.
Since in a relativistic theory a photon which is low-energy
for one (local) observer Alice has different higher energy
for another relatively-boosted observer Bob (also local to the photon)
it must then be the case that photons of any energy are stable if the
relative-locality momentum space is ``DSR compatible", since then evidently
the laws that establish whether or not a photon can decay must be observer
independent.

A key point for my argument is played by the concept of ``decay threshold".
If photons were stable at low energies but could decay at higher energies
there should be a threshold value of the photon energy above which
the decay is allowed. But photons which are below a threshold energy value
for one observer will be above that threshold energy value for other
boosted observers.
A threshold for photon decay could only be the scale $|\ell|^{-1}$
of the relative-locality framework, which can be an invariant
characteristic of the momentum-space geometry.
But if one wants a relative-locality momentum space compatible
with the implementation of any form of (possibly deformed) Lorentz
invariance there cannot be a threshold for photon decay
which is lower than $|\ell|^{-1}$.

Of course, these requirements are perfectly enforced when momentum
space is trivially Minkowski/flat (special relativity),
and indeed there the process $\gamma \rightarrow e^+ e^-$ is strictly
forbidden. One way to see this technically is to derive the formula
that links the opening angle $\theta$
of the (hypothetical) outgoing electron-positron
and the energies $E_+$ and $E_-$
respectively of the positron and electron.
One finds that the process would require $cos \theta >1$ for any $E_+ , E_-$
combination. In particular, this classic special-relativistic result
takes the following form (involving also the electron mass $m$)
\begin{equation}
cos \theta \simeq \frac{2 E_+ E_- + m^2}{2 E_+ E_-
+ m^2 \left( \frac{E_+}{E_-} + \frac{E_-}{E_+} \right) }
\label{pairproductionSR}
\end{equation}
in the limit of ultrarelativistic outgoing particles.

Our stated purpose for this subsection evidently requires
us to reconsider this classic derivation when the
momentum-space is characterized by
(\ref{metricMASTER}), (\ref{connectionMASTER0}), (\ref{connectionMASTERj}),
rather than being trivially Minkowki/flat.

So I start from
\begin{eqnarray}
E_\gamma &=& E_+ + E_- + \beta_1 {\vec{p}}_+ \cdot {\vec{p}}_-
+ \beta_2 E_+ E_-
 \label{phodecay1a}\\
{\vec{p}}_\gamma  &=& {\vec{p}}_+ + {\vec{p}}_-
+ \gamma_{{\!\!\!~}_1} E_+ {\vec{p}}_- + \gamma_{{\!\!\!~}_2} E_- {\vec{p}}_+
\label{phodecay1b}
\end{eqnarray}

From (\ref{phodecay1b}) it follows that
\begin{equation}
p_\gamma^2 = p_+^2 + p_-^2 + 2 p_+ p_- cos \theta +
+ 2 \gamma_{{\!\!\!~}_1} E_+ E_-^2
+ 2 \gamma_{{\!\!\!~}_1} E_+^2 E_- cos\theta
 + 2 \gamma_{{\!\!\!~}_2} E_- E_+^2
 + 2  \gamma_{{\!\!\!~}_2} E_-^2 E_+ cos \theta ~.
\label{phodecay2}
\end{equation}
Here on the right-hand side I already used the restriction
to ultrarelativistic outgoing electron and positron
(but only in terms with factors of $\ell$).
It should be clear that this is where one can look for
pathologies since the deformation of geometry of momentum space
is negligible at small momenta and I am therefore evidently
looking for a possible
ultrarelativistic threshold for photon decay into an electron-positron pair.

From (\ref{phodecay2}), using (\ref{metricMASTER}),
one obtains
\begin{eqnarray}
E_\gamma^2 + (\alpha_1+\alpha_2) E_\gamma^3
 &=& E_+^2 + E_-^2  - 2 m^2
+ \left[ 2 E_+ E_-
+ m^2 \left( \frac{E_+}{E_-} + \frac{E_-}{E_+} \right) \right]
\cos \theta + (\alpha_1+\alpha_2) (E_+^3+E_-^3) + \label{phodecay3}\\
&&+ 2 \gamma_{{\!\!\!~}_1} E_+ E_-^2
 + 2 \gamma_{{\!\!\!~}_2} E_- E_+^2
+ \left[(\alpha_1+\alpha_2)(E_+ + E_-) + 2 \gamma_{{\!\!\!~}_1} E_+
+ 2  \gamma_{{\!\!\!~}_2} E_-\right] E_+ E_- \cos \theta
\nonumber
\end{eqnarray}
For the left-hand side of this result one can use (\ref{phodecay1a}),
which leads to
\begin{eqnarray}
&&E_+^2  + E_-^2 + 2 E_+ E_-
+ 2 (\beta_2 + \beta_1 \cos \theta)(E_+ + E_-) E_+ E_-
+ (\alpha_1+\alpha_2) (E_+ + E_-)^3 =\nonumber\\
&&~~~~~~~~~~~~~~~~~ = E_+^2 + E_-^2  - 2 m^2
+ \left[ 2 E_+ E_-
+ m^2 \left( \frac{E_+}{E_-} + \frac{E_-}{E_+} \right) \right]
\cos \theta + (\alpha_1+\alpha_2) (E_+^3+E_-^3) + \label{phodecay4}\\
&&
~~~~~~~~~~~~~~~~~~~ + 2 \gamma_{{\!\!\!~}_1} E_+ E_-^2
 + 2 \gamma_{{\!\!\!~}_2} E_- E_+^2
+ \left[(\alpha_1+\alpha_2)(E_+ + E_-) + 2 \gamma_{{\!\!\!~}_1} E_+
+ 2  \gamma_{{\!\!\!~}_2} E_-\right] E_+ E_- \cos \theta
\nonumber
\end{eqnarray}

Then taking into account that I am working at leading order in $\ell$,
and assuming again that the outgoing particles
are ultrarelativistic (so that $\ell E_{+} \neq 0$
and $\ell E_{-} \neq 0$
but $\ell m \simeq 0$ and $\ell^2 E_{+}^2 \simeq 0 \simeq \ell^2 E_{+}^2$)
one gets
\begin{equation}
cos \theta \simeq \frac{2 E_+ E_- + m^2+ 2
(\alpha_1
+ \alpha_2 + \beta_1 + \beta_2
- \gamma_{{\!\!\!~}_1} - \gamma_{{\!\!\!~}_2}  )
(E_+ + E_-) E_+ E_- }{2 E_+ E_-
+ m^2 \left( \frac{E_+}{E_-} + \frac{E_-}{E_+} \right) }
\label{pairproductionPRL}
\end{equation}

So I have established that the ``no-photon-decay-switch-on constraint"
is
\begin{eqnarray}
\alpha_1
+ \alpha_2 + \beta_1 + \beta_2
- \gamma_{{\!\!\!~}_1} - \gamma_{{\!\!\!~}_2} \geq 0
\label{goldenrulepart1}
\end{eqnarray}

If instead $\alpha_1
+ \alpha_2 + \beta_1 + \beta_2
- \gamma_{{\!\!\!~}_1} - \gamma_{{\!\!\!~}_2} < 0$
then there would be some high values of $E_+,E_-$
(and correspondingly $E_\gamma$)
for which a real $\theta$ could solve (\ref{pairproductionPRL})
({\it i.e.} such that $|\cos \theta| \leq 1$),
and this in turn would mean that photon decay is allowed at
those high energies.\\
It must be noticed that the energies needed for this are ultra-high
but below-Planckian: assume
for example $\alpha_1
+ \alpha_2 + \beta_1 + \beta_2
- \gamma_{{\!\!\!~}_1} - \gamma_{{\!\!\!~}_2} = - \ell$
then photon decay starts to be allowed already at scales
roughly of order $(m^2 |\ell|^{-1})^{1/3}$ (which indeed, for $m$ the electron mass
and $|\ell|^{-1}$ roughly of order the Planck scale, is $\ll |\ell|^{-1}$).

\subsection{No-pair-production-switch-off constraint}
For the second requirement, the ``no-pair-production-switch-off constraint",
the process I use is electron-positron
pair production in photon-photon collisions:
$\gamma + \Gamma \rightarrow e^+ e^-$
(I adopt a convention such that the energy $\epsilon$ of the photon $\gamma$ is
lower than the energy $E$ of the photon $\Gamma$).

This process if evidently allowed in ordinary special relativity,
and it has been established to actually occur in countless experiments,
of course in a limited range of so-far-experimentally-accessible values
of the energies of the two incoming photons. What could ``go wrong" with
this pair-production process in absence of (not even some deformation of)
Lorentz symmetry? To see this let us gradually lower the value of the energy $\epsilon$
of the ``soft" photon, and ask if there are correspondingly high hard-photon
energies compatible with pair production. In special relativity of course
the answer is always yes: for given low energy $\epsilon$ of the photon $\gamma$
one has that pair production
is always allowed if the other photon $\Gamma$ has energy $E \geq m^2/\epsilon$.
The question ``can a photon of energy $\epsilon$ interact with another photon
to produce an electron-positron pair?" always has positive answer,
for all values of $\epsilon$. This is also a necessary consequence of the fact
that two relatively boosted observers attribute different energy
to a given photon: a relativistic description of such a pair of relatively boosted
observers
evidently excludes the possibility of a threshold for ``pair-production switch-off".\\
And it is also evident that such a threshold for pair-production switch-off
must not be present in any theory providing a relativistic description
of pairs of relatively-boosted observers. So in order for a chosen geometry
of momentum space to be DSR-compatible
it must always be possible for a photon of any energy $\epsilon$ to produce
electron-positron pairs in interactions with at least some sufficiently high-energy photons.
I shall now show that this imposes another nontrivial requirement
on our parameters $\alpha_1,\alpha_2,\beta_1,\beta_2,\gamma_1,\gamma_2$.

The process $\gamma + \Gamma \rightarrow e^+ e^-$
involves one more particle than the photon-decay process considered in the
previous subsection, but the analysis is simplified by the fact that
the constraint I am looking for can be established focusing on
collinear processes.
In fact, the energy $E$ of photons eligible to
produce pairs with a photon of given energy $\epsilon$
is inevitably going to be greater than a certain minimum
value of $E$, which I shall denote with $E_{min}$, for which the process is collinear.
So I start from
\begin{eqnarray}
E_{min} +\epsilon &=& E_+ + E_- + \beta_1 p_+ p_-
+ \beta_2 E_+ E_-
 \label{pairprod1a}\\
p_{min} - \epsilon &=& p_+ + p_- + \gamma_{{\!\!\!~}_1} E_+ p_- + \gamma_{{\!\!\!~}_2} E_- p_+
\label{pairprod1b}
\end{eqnarray}
where I chose to focus on cases where $\epsilon \ll E_{min}$
so that $\ell \epsilon \simeq 0$ even though $\ell E_{min} \neq 0$.
Then one can easily combine (\ref{pairprod1b}) and (\ref{metricMASTER}), also relying on
some of the approximations already exploited in the previous
subsection, to establish
that
\begin{eqnarray}
E_{min} + \frac{\alpha_1 + \alpha_2}{2} E_{min}^2
- \epsilon =
 E_+ + E_- -\frac{m^2}{2 E_+}-\frac{m^2}{2 E_-}
 + \frac{\alpha_1 + \alpha_2}{2} (E_+^2 + E_-^2)
+ (\gamma_{{\!\!\!~}_1} + \gamma_{{\!\!\!~}_2}) E_+ E_- ~.
\label{pairprod2}
\end{eqnarray}

A further simplification is obtained by noticing that
the zero-th order ($\ell \rightarrow 0$) solution is such that $E_+ = E_- =E_{min}/2$,
and that this zero-th order property can be safely assumed, consistently
with the approximations I am adopting, to still hold within terms with already
small prefactors of $m^2$ ($\ll E_{min}^2$) or $\ell$ ($\ll 1/E_{min}$).
So from (\ref{pairprod2}) one has
\begin{eqnarray}
E_{min} + \frac{\alpha_1 + \alpha_2}{2} E_{min}^2
- \epsilon =
 E_+ + E_- - 2 \frac{m^2}{E_{min}}
 + \frac{\alpha_1 + \alpha_2}{4} E_{min}^2
+ (\gamma_{{\!\!\!~}_1} + \gamma_{{\!\!\!~}_2}) \frac{E_{min}^2}{4}
\label{pairprod3}
\end{eqnarray}
and from (\ref{pairprod1a}) one has
\begin{eqnarray}
E_{min} +\epsilon &=& E_+ + E_- + (\beta_1 + \beta_2) \frac{E_{min}^2}{4}
\label{pairprod4}
\end{eqnarray}

Combining (\ref{pairprod3}) and (\ref{pairprod4}) one finds
\begin{eqnarray}
2 \epsilon =
2 \frac{m^2}{E_{min}}
+  (\alpha_1 + \alpha_2) \frac{E_{min}^2}{4}
+ (\beta_1 + \beta_2) \frac{E_{min}^2}{4}
- (\gamma_{{\!\!\!~}_1} + \gamma_{{\!\!\!~}_2}) \frac{E_{min}^2}{4}
\label{pairprod5}
\end{eqnarray}
So in summary $E_{min}$ must satisfy the condition
\begin{eqnarray}
E_{min} - (\alpha_1 + \alpha_2 + \beta_1 + \beta_2 - \gamma_{{\!\!\!~}_1} - \gamma_{{\!\!\!~}_2}) \frac{E_{min}^3}{4\epsilon}
 = \frac{m^2}{\epsilon}
\label{pairprod5}
\end{eqnarray}
This allows us to conclude that in order to avoid
the ``pair-production switch-off" one must enforce
\begin{eqnarray}
\alpha_1
+ \alpha_2 + \beta_1 + \beta_2
- \gamma_{{\!\!\!~}_1} - \gamma_{{\!\!\!~}_2} \leq 0
\label{goldenrulepart2}
\end{eqnarray}
If instead $\alpha_1
+ \alpha_2 + \beta_1 + \beta_2
- \gamma_{{\!\!\!~}_1} - \gamma_{{\!\!\!~}_2} > 0$
then one could find values of $\epsilon$ small enough (values of $m^2/\epsilon$
large enough) that (\ref{pairprod5}) would admit no solution, so that
indeed pair-production would be switched off.\\
And once again it turns out
that the issue is not confined to the ``Planckian regime":
for example if $\epsilon$ is $\sim 10^{-5}eV$ in standard special relativity
(flat/Minkowski momentum-space geometry) pair production can occur whenever
the other photon has energy $\geq 3 \cdot 10^{17}eV$, whereas with
the deformation scheme I am considering, if for example $\alpha_1
+ \alpha_2 + \beta_1 + \beta_2
- \gamma_{{\!\!\!~}_1} - \gamma_{{\!\!\!~}_2} \simeq \ell$,
for $\epsilon \sim 10^{-5}eV$ the pair-production process is already
switched off: if $\epsilon \sim 10^{-5}eV$ then according to (\ref{pairprod5})
pair production cannot occur for any value of $E$,
if $m$ is the
electron mass, $|\ell|^{-1}$ is roughly of order the Planck scale,
and  $\alpha_1
+ \alpha_2 + \beta_1 + \beta_2
- \gamma_{{\!\!\!~}_1} - \gamma_{{\!\!\!~}_2} \simeq \ell$.

\subsection{A golden rule}
Combining the results derived in the previous two subsections,
summarized in Eq.~(\ref{goldenrulepart1}) and Eq.~(\ref{goldenrulepart2}),
I obtain the following constraint on the deformation parameters:
\begin{eqnarray}
\alpha_1
+ \alpha_2 + \beta_1 + \beta_2
- \gamma_{{\!\!\!~}_1} - \gamma_{{\!\!\!~}_2} = 0
\label{goldenrule}
\end{eqnarray}
Following the logical line of the observations I reported in this section
one concludes that for such geometries of relative-locality momentum spaces\\
\indent IF the geometry of the momentum space violates this constraint
 ({\it i.e.} IF $\alpha_1
+ \alpha_2 + \beta_1 + \beta_2
- \gamma_{{\!\!\!~}_1} - \gamma_{{\!\!\!~}_2} \neq 0$)\\
\indent THEN theories on such relative-locality momentum spaces require the
introduction of a preferred ``ether" frame (the laws of physics on such relative-locality momentum spaces
are observer-dependent).\\
The condition (\ref{goldenrule}) is {\underline{necessary}} for a relative-locality
momentum space to be DSR-compatible. But is (\ref{goldenrule}) also
a {\underline{sufficient}} condition for a relative-locality momentum space to be DSR-compatible?\\
That is, rewriting more explicitly,
are we assured that on relative-locality momentum spaces that satisfy (\ref{goldenrule})
it will be possible to introduce laws that involve the scale $\ell$ characteristic
of the momentum-space geometry and yet are observer-independent laws?\\
The derivations I have offered so far only establish (\ref{goldenrule}) as
a {\underline{necessary}} condition for DSR-compatibility, but
in the following sections, by investigating a few examples of relative-locality momentum spaces,
 I provide evidence
that when  (\ref{goldenrule})  is  satisfied one does manage
to base on such relative-locality momentum spaces laws of physics which are observer independent.\\
So I {\underline{conjecture}} that (\ref{goldenrule}) is also sufficient for DSR-compatibility.\\
If there are any counter-examples to my conjecture of sufficiency of (\ref{goldenrule}) 
for DSR compatibility I expect them
to be somewhat ``pathological" in one way or another: {\it a posteriori}
I can recognize in Eq.~(\ref{goldenrule})
a compact summary, a sort of ``golden rule", reflecting 
a significant part of my experience working for
more than a decade with scenarios of Lorentz-symmetry deformation.
And even just the established necessity for DSR compatibility of (\ref{goldenrule})
qualifies it as a ``golden rule" for the phenomenology
of deformations of Lorentz symmetry centered on momentum space:
 (\ref{goldenrule}) allows one to quickly conclude that
a certain geometry of momentum space does not admit observer-independent laws of physics,
whereas reaching the same conclusion using more formal tools of investigation
 can be an endless task.

For the illustrative example of parametrization of momentum space on which I focused
my ``criteria" (necessary conditions) for DSR compatibility took the shape of 
the simple algorithmic requirement (\ref{goldenrule}).
I expect that for most other parametrizations of the momentum-space geometry
my criteria will again produce simple algorithmic recipes. An exception to this
expectation of simple algorithmic implementation of the criteria may perhaps
be found in scenarios such as the one in Ref.~\cite{dsrgzk}, where the deformation is
not structured simply as a power series in the components of momentum (in particular
Ref.~\cite{dsrgzk} tentatively contemplates
deformations involving powers of components of momentum divided by powers 
of the mass of the particle): in such cases the region of energy scales of interest
in physics applications
can be outside the region of applicability of the leading-order approximation
and, while my criteria would still apply, the algorithmic formulation of my criteria
might be unavailable.

\section{Some examples with torsion-free momentum space}\label{dsr1sec}
The remainder of this manuscript has two goals:\\
(i) test the reliability of the ``golden rule" derived in the
previous section, adopted tentatively as a (not only necessary but also)
sufficient condition for DSR-compatibility,
by looking at momentum spaces where the golden rule
is satisfied and checking explicitly that some formulation of deformed
Lorentz symmetry is available;\\
(ii) illustrate some of the peculiarities that can
characterize deformed Lorentz
symmetry on a relative-locality momentum space.

The examples of relative-locality momentum spaces I consider in this section
and the next section
are characterized by being setups which were already of interest during the earliest
stages of DSR research\footnote{The DSR proposal was not exclusively intended~\cite{dsr1Edsr2}
 for allowing an observer-independence/relativistic description of deformed laws on momentum space.
Examples of other options that were and are considered include introducing the second
observer-independent scale primitively in spacetime structure or primitively at the
level of the (deformed) de Broglie relation between wavelength and momentum~\cite{dsr1Edsr2,gacdsrrev2010}.
However, since I am here focusing on the relative-locality framework, with its assumption
that momentum space is primitive~\cite{prl,grf2nd}, the DSR-research tools I shall need all
come from previous attempts to construct a DSR framework centered on nonlinear laws
on momentum space.}.
Specifically in this section I start with two examples of torsionless  momentum spaces,
with in particular one example matching exactly the form of  nonlinear laws in momentum
space that were considered already in Ref.~\cite{dsr1Edsr2}, as an attempt to illustrate
the idea that one might try to have a relativistic theory with an observer-independent
length/inverse-momentum scale, while preserving the observer-independence of the laws of
physics and the overall relativistic nature of the theory. Since relative-locality momentum
spaces are always flat in leading order~\cite{prl,grf2nd} (the lowest-order contribution
to curvature of the metric on a relative-locality
momentum space is of order $\ell^2$), and these are also torsionless cases, these are
evidently not the most interesting scenarios from the novel relative-locality perspective,
but I find nonetheless somewhat reassuring that indeed the golden rule does perform
well in my two examples of this sort.\\
Then in the next section I consider two other examples of a traditional type
in the DSR literature, examples largely inspired by
properties of the $\kappa$-Poincar\'e Hopf algebra~\cite{lukieIW,majidruegg,kpoinap}.
The $\kappa$-Poincar\'e-inspired nonlinearities for momentum-space laws appeared
immediately~\cite{dsr1Edsr2} as a natural candidate for obtaining a DSR-compatible framework.
But this $\kappa$-Poincar\'e opportunity for DSR research has remained in
a ``sub judice status" mainly as a result that, as already noticed in Ref.~\cite{dsr1Edsr2},
some of the attempts to build quantum field theories with $\kappa$-Poincar\'e
structures appeared not to enforce the observer independence of the laws of physics.\\
The observations I here report in the next section on the $\kappa$-Poincar\'e-inspired
relative-locality momentum space (actually two versions of it) go some way in the direction
of establishing more firmly the case for DSR compatibility, and do so while providing
further evidence of robustness of the use of my ``golden rule" as a sufficient condition
for DSR-compatibility.\\
Within the relative-locality framework these $\kappa$-Poincar\'e-inspired
relative-locality momentum spaces are representatives of the case of torsionful momentum spaces,
and are therefore rather nontrivial.

As announced, here and in the next section
I work again exclusively in leading order in $\ell$.
And I work in $1+1$ momentum-space dimensions, so that the formulas
take their simplest possible form and it will be easier to highlight
the conceptual steps.

\subsection{Brief summary of previous results on the ``DSR1" }

As announced, the first definite example
I consider of relative-locality momentum space
is actually a setup that was already considered in the DSR literature,
and there known as ``DSR1", often used (already in Ref.~\cite{dsr1Edsr2})
as a way to illustrate some of the ingredients that could be used when attempting
to produce a DSR framework, with special-relativistic laws deformed by
a length/invere-momentum scale which are however relativistic/observer-independent laws.
A key issue of interest in the DSR literature is indeed the compatibility
of a given choice of on-shell relation and a given choice
of momentum-conservation laws with (possibly deformed) Lorentz invariance.

In this DSR1 setup the on-shell relation takes the form~\cite{dsr1Edsr2}
\begin{equation}
m^2 = p_0^2 - p_1^2 + \ell p_0 p_1^2 ~.
\label{metricdsr1}
\end{equation}
which is invariant under the DSR1-deformed boost action~\cite{dsr1Edsr2}
\begin{equation}
[N, p_0] =  p_1 ~,~~~
[N, p_1] =  p_0 + \ell p_0^2 + \frac{\ell}{2} p_1^2
\label{boosts}
\end{equation}
And an example of DSR1-deformed law of conservation of momentum
that would be compatible with the boost action (\ref{boosts}) is~\cite{dsr1Edsr2}:
\begin{eqnarray}
p^{(a)}_0 + p^{(b)}_0 + \ell p^{(a)}_1  p^{(b)}_1 &=&
p^{(c)}_0 + p^{(d)}_0 + \ell p^{(c)}_1  p^{(d)}_1
 \label{dsr1econs}\\
p^{(a)}_1 + p^{(b)}_1 + \ell p^{(a)}_0  p^{(b)}_1 + \ell p^{(a)}_1  p^{(b)}_0 &=&
p^{(c)}_1 + p^{(d)}_1 + \ell p^{(c)}_0  p^{(d)}_1 + \ell p^{(c)}_1  p^{(d)}_0
\label{dsr1pcons}
\end{eqnarray}
for an event $a + b \rightarrow c+d$.

The on-shell relation of this DSR1 model, (\ref{metricdsr1}),
is evidently of a type that can emerge from a choice of metric on momentum space.
But I must still comment on the law of composition of momenta,
which according to Refs.~\cite{prl,grf2nd} codifies the affine connection
on momentum space.
It is nearly self-evident that the conservation laws  (\ref{dsr1econs})-(\ref{dsr1pcons})
implicitly use the composition law
\begin{equation}
(k \oplus p)_0 = k_0 + p_0 + \ell k_1 p_1~,~~~(k \oplus p)_1 = k_1 + p_1
+ \ell k_0 p_1 + \ell p_0 k_1~,
\label{connectiondsr1}
\end{equation}
which is commutative and therefore,
according to the criteria introduced in Refs.~\cite{prl,grf2nd},
corresponds to a torsionless connection\footnote{Moreover, this DSR1 setup, besides being torsionless,
 can also be mapped by a diffeomorphism onto the flat momentum space of special relativity.
Some authors (see, {\it e.g.}, Refs.~\cite{iranNODSR1,ahluNODSR1})
have stressed that if the overall
theoretical framework introduced on momentum space is diffeomorphism invariant
then of course the DSR1 kinematics in such instances would be just reproducing the
physical predictions of ordinary special relativity.
Not denying the potential appeal of insisting on such a ``momentum-space general covariance",
I remain interested also in the possibility that the laws of physics on momentum space
may not be diffeomorphism invariant, in which case setups such as this DSR1 kinematics
and the other setup discussed in the next subsection could be highly nontrivial.
Readers who are only interested in setting up theories whose overall structure
ensures momentum-space diffeomorphism invariance  may well skip this section
and proceed to the next Section~\ref{torsionsec}.}.\\
In order to verify explicitly that the composition law (\ref{connectiondsr1})
corresponds to the conservation law (\ref{dsr1econs})-(\ref{dsr1pcons})
I must  check
 that (\ref{dsr1econs})-(\ref{dsr1pcons})
 is equivalent to
$$(((p^{(a)} \oplus p^{(b)} ) \oplus p^{(c)}) \oplus p^{(d)})_\mu = 0~,$$
which in light of (\ref{connectiondsr1}) can be rewritten as
\begin{eqnarray}
\!\!\!\!\!\!\!\!\!\!\!\!\!\!\!\!\!\!\!\!\!\!\!\!\!\!\!\!\!\!\!\!\!\!\!\!\!\!\!\!\!\!\!\!\!
\!\!\!\!\!\!\!\!\!\!\!\!\!\!\!\!\!\!\!\!\!
p^{(a)}_0 + p^{(b)}_0 + p^{(c)}_0 + p^{(d)}_0
+\ell p^{(a)}_1  p^{(b)}_1
+\ell p^{(a)}_1  p^{(c)}_1
+\ell p^{(a)}_1  p^{(d)}_1
+\ell p^{(b)}_1  p^{(c)}_1
+\ell p^{(b)}_1  p^{(d)}_1
+ \ell p^{(c)}_1  p^{(d)}_1
=0~,
 \nonumber\\
\!\!\!\!\!\!\!\!\!\!\!\!\!\!\!\!\!\!\!\!
\!\!\!\!\!\!\!\!\!\!\!\!\!\!\!\!\!\!\!\!\!\!\!\!
 p^{(a)}_1 \! + \! p^{(b)}_1 \! + \! p^{(c)}_1 \! + \! p^{(d)}_1
 \! + \! \ell [p^{(a)}_0  p^{(b)}_1 \!\! + \! p^{(a)}_1  p^{(b)}_0
 \!\! + \! p^{(a)}_0  p^{(c)}_1 \!\! + \! p^{(a)}_1  p^{(c)}_0 \! +
 ~~~~~~~~~~~~~~~~~~~~~~~~~ ~~~~~~~~~~~~~ \nonumber\\
 \!\! + \! p^{(a)}_0  p^{(d)}_1 \!\! +  \! p^{(a)}_1  p^{(d)}_0
 \!\! + \! p^{(b)}_0  p^{(c)}_1 \!\! + \! p^{(b)}_1  p^{(c)}_0
 \!\! + \! p^{(b)}_0  p^{(d)}_1 \!\! + \! p^{(b)}_1  p^{(d)}_0
 \!\! + \! p^{(c)}_0  p^{(d)}_1 \!\! + \! p^{(c)}_1  p^{(d)}_0
=0
\nonumber
\end{eqnarray}
The specific case of (\ref{dsr1econs})-(\ref{dsr1pcons})
is actually obtained for
$$(((p^{(a)} \oplus p^{(b)} ) \oplus
 [\ominus p^{(c)})] \oplus [\ominus p^{(d)}])_\mu = 0~,$$
 which may be viewed as the case of two incoming and two outgoing
 momenta~\cite{prl},
where $\ominus$ is the antipode of the $\oplus$ in (\ref{connectiondsr1})
\begin{equation}
(\ominus p)_0 = - p_0 + \ell p_1 p_1~,~~~
(\ominus p)_1 = - p_1 + 2 \ell p_0 p_1  ~.
\label{antipode}
\end{equation}
Indeed one easily checks that $p \oplus (\ominus p) =0$, as required for the antipode.

This observation finds confirmation in the fact that
\begin{eqnarray}
(((p^{(a)} \oplus p^{(b)} ) \oplus
 [\ominus p^{(c)})] \oplus [\ominus p^{(d)}])_0 = 0
 & \Longleftrightarrow  & p^{(a)}_0 + p^{(b)}_0 + \ell p^{(a)}_1  p^{(b)}_1 =
p^{(c)}_0 + p^{(d)}_0 + \ell p^{(c)}_1  p^{(d)}_1
 \nonumber\\
(((p^{(a)} \oplus p^{(b)} ) \oplus
 [\ominus p^{(c)})] \oplus [\ominus p^{(d)}])_1 = 0
 & \Longleftrightarrow &
 p^{(a)}_1 + p^{(b)}_1 + \ell p^{(a)}_0  p^{(b)}_1 + \ell p^{(a)}_1  p^{(b)}_0 =
p^{(c)}_1 + p^{(d)}_1 + \ell p^{(c)}_0  p^{(d)}_1 + \ell p^{(c)}_1  p^{(d)}_0
\nonumber
\end{eqnarray}
where the equivalences are established of course dropping terms which
vanish when the conservation laws are enforced (and taking into
account that I am working in leading order in $\ell$).

So it is at this point fully established
that the geometry of momentum space that matches the prescriptions
of the DSR1 model is characterized in terms of
the on-shell relation (\ref{metricdsr1})
and the composition law
(\ref{connectiondsr1}).

With this observation we are ready to
 check that the DSR1 setup is consistent with the ``golden rule".
For that purpose
let me observe that, using the parametrization of the previous section,
the DSR1 on-shell relation and composition law
correspond to the choice of parameters $\alpha_1 = \ell$, $\alpha_2 = 0$,
$\beta_1 = \ell$, $\beta_2 = 0$,
$\gamma_1 = \gamma_2 = \ell$, so that the golden rule
is satisfied: $\alpha_1
+ \alpha_2 + \beta_1 + \beta_2
- \gamma_{{\!\!\!~}_1} - \gamma_{{\!\!\!~}_2} = 0$.
And indeed it is easy to verify~\cite{dsr1Edsr2} that when the
conservation laws (\ref{dsr1econs})-(\ref{dsr1pcons})
hold for one observer, say Alice, they also hold for any observer
boosted according to Eq.~(\ref{boosts}) with respect to Alice.\\
We can therefore recognize in the DSR1 model
a first success for the possibility of viewing the ``golden rule" as a sufficient
condition for DSR-compatibility.

\subsection{Another torsionless example, with abelian
energy composition}
Let me now consider a slightly different setup, which was not previously
discussed in the literature: a setup similar to the one
of the previous subsection (in particular with torsionless momentum space)
and with the added ``simplicity bonus" of undeformed additivity\footnote{Since
my focus is on relativistic properties, postponing to future studies
a more detailed discussion of physics applications, I shall not
here dwell on the advantages of having undeformed additivity
of energy. This is however an issue that was discussed in some detail
already in the doubly-special-relativity literature, as seen
for example in Ref.~\cite{gianlucaenergy} and references therein.}
of energy (but not of spatial momenta).

For this second example of torsionless momentum space
I take a  metric such that the one-shell condition is
\begin{equation}
m^2 = p_0^2 - p_1^2 + 2 \ell p_0 p_1^2
\label{metric}
\end{equation}
and for the affine connection I take one such that
\begin{equation}
(k \oplus p)_1 = k_1 + p_1 + \ell k_0 p_1 + \ell p_0 k_1 ~,~~~
(k \oplus p)_0 = k_0 + p_0~,
\label{connection}
\end{equation}
which is indeed commutative, as required for
a torsionless momentum space in the sense of Refs.~\cite{prl,grf2nd}.

I note down the antipode that follows from (\ref{connection})
\begin{equation}
(\ominus p)_1 = - p_1 + 2 \ell p_0 p_1  ~,~~~
(\ominus p)_0 = - p_0
\label{antipode}
\end{equation}
And I also observe that from (\ref{connection}) it follows that
\begin{equation}
[(k \oplus p) \oplus q]_1 = k_1 + p_1 + q_1 + \ell k_0 (p_1 + q_1)
+ \ell p_0 (k_1 + q_1) + \ell q_0 (k_1 + p_1) ~,~~~
[(k \oplus p) \oplus q]_0 = k_0 + p_0 + q_0
\label{tribody}
\end{equation}
For this other torsionless
momentum space that I want to consider
I have already laid out the ingredients needed for applying my ``golden rule".
From the on-shell relation (\ref{metric})
and the composition law (\ref{connection}) we see that, using the parametrization
of the previous section, this is a case
with $\alpha_1 = 2 \ell$, $\alpha_2 = 0$,
$\beta_1 = 0$, $\beta_2 = 0$,
$\gamma_1 = \gamma_2 = \ell$, and therefore
we have here another case in which
the golden rule is satisfied: $\alpha_1
+ \alpha_2 + \beta_1 + \beta_2
- \gamma_{{\!\!\!~}_1} - \gamma_{{\!\!\!~}_2} = 0$.\\
So there is no {\it a priori} reason to expect that such a setup would require
a ``preferred-frame formulation" (since the necessary condition for DSR-compatibility is satisfied),
and if one assumes that the ``golden rule" is also a sufficient condition for DSR-compatibility
one must expect that
it will be possible to introduce in this setup a satisfactory notion
of deformed Lorentz symmetry. Indeed this is the case, as I shall now
easily show.\\
Specifically I shall verify
that a satisfactory
description of deformed Lorentz symmetry of this torsionless
momentum space is obtained
in terms of the following boost generator
\begin{equation}
[N, p_0] =  p_1 - \ell p_0 p_1 ~,~~~
[N, p_1] =  p_0 + \ell p_0^2 + \ell p_1^2~.
\label{boostsnew}
\end{equation}
I start by observing that
the on-shell relation is invariant:
\begin{equation}
[N, p_0^2 - p_1^2 + 2 \ell p_0 p_1^2 ] = 0~.
\label{invariantshell}
\end{equation}
And it is only slightly more tedious to check that the boost
(\ref{boostsnew})
ensures the covariance of the laws of conservation
of momentum based on the composition law (\ref{connection}).
I start checking this covariance for the conservation law
\begin{equation}
(k \oplus p) = 0
\label{twobodyzero}
\end{equation}
Acting with the boost $N$ one finds that
\begin{equation}
[N, (k \oplus p)_0]=[N, (k_0 + p_0)]= k_1 - \ell k_0 k_1 +p_1 - \ell p_0 p_1 =
k_1 + p_1 + \ell p_0 k_1 +  \ell k_0 p_1 - \ell (p_0 + k_0) ( p_1 + k_1)= 0
\label{covatwotime}
\end{equation}
where on the right-hand side I of course
used the conservation law itself (and took into account that I am working
in leading order in $\ell$). And similarly one finds
that
\begin{eqnarray}
[N, (k \oplus p)_1]=[N, (k_1 + p_1 + \ell k_0 p_1 + \ell p_0 k_1)]=
 k_0+p_0 +\ell p_0^2 +\ell k_0^2 + \ell p_1^2 + \ell k_1^2 +
2 \ell k_1 p_1 + 2 \ell k_0 p_0 \nonumber\\
~~~~~~~~
=  k_0+p_0 +\ell  (p_0 + k_0)^2 + \ell (p_1 + k_1)^2 = 0
~~~~~~~~
\label{covatwospace}
\end{eqnarray}
where again on the right-hand side I
used the conservation law itself.\\
The covariance of the conservation law $(k \oplus p) = 0$ under
the  boost  (\ref{boostsnew})
is evidently confirmed by (\ref{covatwotime}) and (\ref{covatwospace}).

Let me also check explicitly
the covariance of the conservation law
\begin{equation}
(k \oplus p) \oplus q = 0
\label{tribodyzero}
\end{equation}
[The covariance of the 4-particle conservation, $[(k \oplus p) \oplus q] \oplus r = 0$,
and of the general $N$-particle conservation is easily checked analogously.]

Acting with the boost $N$ on the $0$-component of (\ref{tribodyzero}) one finds
that
\begin{eqnarray}
&& \!\!\!\!\!\! [N, [(k \oplus p) \oplus q]_0]=[N, (k_0 + p_0 + q_0)]=
k_1 - \ell k_0 k_1 +p_1 - \ell p_0 p_1 +q_1 - \ell q_0 q_1 = \nonumber\\
&& ~~~~~~~~ k_1 + p_1 +q_1 + \ell p_0 k_1 +  \ell k_0 p_1
+ \ell q_0 k_1 +  \ell k_0 q_1
+ \ell p_0 q_1 +  \ell q_0 p_1
 - \ell (p_0 + k_0 +  q_0 ) ( p_1 + k_1 + q_1)= 0
\label{covatritime}
\end{eqnarray}
 And similarly for the $1$-component one finds
that
\begin{eqnarray}
[N, ((k \oplus p) \oplus q)_1]=[N,
(k_1 + p_1 + q_1 + \ell k_0 (p_1 + q_1)
+ \ell p_0 (k_1 + q_1) + \ell q_0 (k_1 + p_1))]=
~~~~~~~~~~~~~~~~~~~~~~~~~~~~~~~~~~~~~~~~~~~ \nonumber\\
~~~~~~ = k_0+p_0 +q_0 +\ell p_0^2 +\ell k_0^2 +\ell q_0^2  + \ell p_1^2 + \ell k_1^2
+\ell q_1^2 + 2 \ell k_1 p_1 + 2 \ell k_0 p_0
+ 2 \ell k_1 q_1 + 2 \ell k_0 q_0
+ 2 \ell p_1 q_1 + 2 \ell p_0 q_0\nonumber\\
= k_0+p_0 +q_0 +\ell  (p_0 + k_0 + q_0)^2 + \ell (p_1 + k_1 +q_1)^2 = 0
~~~~~~~~~~~~~~~~~~~~~~~~~~~~~~~~~~~~~~~~~~~~~~~~~~~~~
\label{covatrispace}
\end{eqnarray}
The covariance of the conservation law $(k \oplus p) \oplus q = 0$ under the
boost (\ref{boostsnew})
is evidently confirmed by (\ref{covatritime}) and (\ref{covatrispace}).

So the assumption that golden rule should be a sufficient condition for DSR-compatibility
was again successful: also for the momentum space considered
in this subsection the golden rule is satisfied and a satisfactory
description of deformed
Lorentz symmetry was found.

\section{Deformed Lorentz symmetry on momentum spaces with torsion}\label{torsionsec}
As announced, my next task is to verify the efficacy of the golden rule (and
give explicit examples of boost transformations that implement
deformed Lorentz symmetry) in cases in which the momentum space
has torsion, and specifically for some cases of $\kappa$-Poincar\'e-inspired momentum
spaces. The associated nonlinear laws on momentum space are cases that were immediately perceived~\cite{dsr1Edsr2,jurekdsr1} as promising candidates for the construction
of a DSR-compatible framework, even though, as mentioned,
they have remained to some extent ``sub judice". Some of the observations I here report,
besides providing additional tests of the robustness of my ``golden rule",
could be relevant for building a more robust case for
considering $\kappa$-Poincar\'e-inspired nonlinearities on momentum space
as a viable candidate for the construction of a DSR framework.

\subsection{Deformed Lorentz symmetry on a momentum space with torsion
and modified dispersion}\label{kappauno}
My first example of momentum space with torsion is the ``$\kappa$-momentum space",
which, as stressed in Ref.~\cite{anatomy}, may deserve special
interest from the relative-locality perspective of Refs.~\cite{prl,grf2nd}
since it is inspired by some properties of the
much-studied $\kappa$-Poincar\'e Hopf algebra~\cite{lukieIW,majidruegg,kpoinap}.

From the viewpoint of contributing to the development of the relative-locality
framework of Refs.~\cite{prl,grf2nd}, the study of deformed boost transformations on
this momentum space is particularly significant since this is the only
relative-locality momentum space for which
a solid relativistic description of distant observers at rest was explicitly obtained~\cite{anatomy}.

Following Ref.~\cite{anatomy} I
characterize $\kappa$-momentum space through a
 momentum-space metric such that the on-shell (``dispersion") relation is
\begin{equation}
m^2 = p_0^2 - p_j^2 + \ell p_0 p_j^2
\label{metrictorsy}
\end{equation}
and the torsionful momentum-space affine connection is
such that
\begin{equation}
(k \oplus p)_j = k_j + p_j + \ell k_0 p_j  ~,~~~
(k \oplus p)_0 = k_0 + p_0~.
\label{connectiontorsy}
\end{equation}
This composition law can be qualified as ``Majid-Ruegg composition law"~\cite{anatomy}
(with the associated ``Majid-Ruegg connection" on momentum space)
since it is primarily inspired by an approach to the description
of the $\kappa$-Poincar\'e Hopf algebra
first introduced by Majid and Ruegg in Ref.~\cite{majidruegg}.
I shall compactly refer to it as the ``MR composition law".

I note down the antipode for the MR composition law:
\begin{equation}
(\ominus p)_j = - p_j + \ell p_0 p_j  ~,~~~
(\ominus p)_0 = - p_0~,
\label{antipodetorsy}
\end{equation}
which indeed, as verified by direct application of (\ref{connectiontorsy}),
is such that $p \oplus (\ominus p) = 0$.\\
And I also observe that from (\ref{connectiontorsy}) it follows that
\begin{equation}
[(k \oplus p) \oplus q]_1 = k_1 + p_1 + q_1 +  \ell k_0 p_1
+  \ell k_0 q_1 +  \ell p_0 q_1  ~,~~~
[(k \oplus p) \oplus q]_0 = k_0 + p_0 + q_0
\label{tribodytorsy}
\end{equation}

It is easy to verify that also this $\kappa$-momentum space
fits the demands of the ``golden rule".
In fact, from the on-shell relation (\ref{metrictorsy})
and the composition law (\ref{connectiontorsy}) we see that,
using the parametrization
of the Sec.~\ref{goldenrulesec}, this is a case
with $\alpha_1 = \ell$, $\alpha_2 = 0$,
$\beta_1 = 0$, $\beta_2 = 0$,
$\gamma_1 = \ell$, $\gamma_2 = 0$, and therefore
indeed the golden rule is satisfied: $\alpha_1
+ \alpha_2 + \beta_1 + \beta_2
- \gamma_{{\!\!\!~}_1} - \gamma_{{\!\!\!~}_2} = 0$.\\
One must therefore expect that also on the $\kappa$-momentum space
 a satisfactory notion
of deformed Lorentz symmetry is available, which indeed is what I shall now
 describe.\\
 With respect to the torsionless cases considered in the previous section,
 our torsionful $\kappa$-momentum space is going to require
 a somewhat more sophisticated type of
 description of the action of boosts.
Let me start by introducing the action of a boost on the momentum
of a particle:
\begin{equation}
[N, p_0] =  p_1 ~,~~~
[N, p_1] =  p_0 + \ell p_0^2 + \frac{\ell}{2} p_1^2
\label{booststorsy}
\end{equation}
This prescription indeed ensures the
invariance of the on-shell relation on the $\kappa$-momentum space:
\begin{equation}
[N, p_0^2 - p_1^2 +  \ell p_0 p_1^2] =
2 p_0 p_1 - 2 p_1 (p_0 + \ell p_0^2 + \frac{\ell}{2} p_1^2)
+\ell p_1^3 +2 \ell p_0^2 p_1 = 0
\label{invariantshelltorsy}
\end{equation}
Evidently the torsion of the $\kappa$-momentum space
poses no particular challenge for the description of boosts
on the momentum of a single particle: the torsion characterizes
composition of momenta and is therefore not felt when boosts act
on the momentum of a single particle. The simplicity of the check of invariance
of the on-shell relation, (\ref{invariantshelltorsy}),
confirms this point.\\\
However, torsion inevitably affects how boosts act on
momenta obtained composing two or more single-particle momenta.
Previously, when I examined in Sec.~\ref{dsr1sec} some
 torsionless cases,
I found that it was possible to
simply impose that the boost of a two-particle
event $e_{k \oplus p}$ would be governed by
$$[N_{torsionless},p \oplus k]
= [N_{torsionless}^{(p)}+N_{torsionless}^{(k)},p \oplus k] $$
where on the left-hand side I kept the notation of
a generic boost action, while on the right-hand side I decomposed
the boost into two pieces, each given in terms
of a boost acting exclusively on a certain momentum in the event.
Essentially this means that for the torsionless cases that I considered
in the previous section I found that one could keep the standard
concept of a ``total boost" generator obtained by combining trivially
the boost generators acting on each individual particle.
One can easily retrace the availability of this option to
the fact that in the cases I considered in the previous section
the law of composition
of momenta was symmetric under exchange of the particles.\\
With torsion in momentum space this simplicity is lost:
the lack of symmetry under exchange of particles
of the composition law (\ref{connectiontorsy})
precludes, as
 one can easily verify,
  the possibility of adopting
a ``total boost generator" given by a trivial sum of single-particle
boost generators.
There is no choice of $N^{(p)}$ capable of ensuring
that $[N^{(p)}+N^{(k)}+N^{(q)},((k \oplus p) \oplus q)_\mu] $ vanishes
whenever $((k \oplus p) \oplus q)_\mu =0$.

What does work on the torsionful $\kappa$-momentum space,
as I shall show, is adopting
$$N^{(k \oplus p)} =N^{(k)}+N^{(p)} + \ell k_0 N^{(p)}$$
and accordingly
$$N^{((k \oplus p) \oplus q)}
=N^{(k)}+N^{(p)}+N^{(q)} + \ell k_0 N^{(p)} + \ell k_0 N^{(q)}
+\ell p_0 N^{(q)}$$

Let us verify that indeed these boost actions
ensure compatibility with the
conservation laws obtained from the MR composition law.
For the $0$-th component of $k \oplus p = 0$ one finds
\begin{equation}
[N^{(k)}+N^{(p)} + \ell k_0 N^{(p)}, (k \oplus p)_0]=
[N^{(k)}+N^{(p)} + \ell k_0 N^{(p)}, k_0 + p_0]= k_1+p_1 + \ell k_0 p_1= 0
\label{covatwotimetorsy}
\end{equation}
where on the right-hand side I of course
used the conservation law itself.\\
Similarly for the $1$-component of $k \oplus p = 0$
it turns out that
\begin{eqnarray}
[N^{(k)}+N^{(p)} + \ell k_0 N^{(p)}, (k \oplus p)_1]
 &=& [N^{(k)}+N^{(p)} + \ell k_0 N^{(p)}, k_1 + p_1 + \ell k_0 p_1]=\nonumber\\
 &= & k_0+\ell k_0^2 + \frac{\ell}{2} k_1^2+ p_0+\ell p_0^2  + \frac{\ell}{2} p_1^2
  + \ell k_1 p_1 + 2 \ell k_0 p_0 \nonumber\\
 &= & k_0+p_0 +\ell  (p_0 + k_0)^2 + \frac{\ell}{2} (p_1 + k_1)^2 = 0
~~~~~~~~
\label{covatwospacetorsy}
\end{eqnarray}
where again on the right-hand side I
used the conservation law $(k \oplus p)_\mu = 0$ itself
(and took again into account that
I am working at leading order in $\ell$).

So the description of boosts given by $N^{(k)}+N^{(p)} + \ell k_0 N^{(p)}$
does ensure the relativistic covariance of the conservation law $k \oplus p = 0$.\\
Let me also check that the corresponding description of boosts
on momenta composed
of 3 single-particle momenta, $N^{(k)}+N^{(p)}+N^{(q)} + \ell k_0 N^{(p)}
+ \ell k_0 N^{(q)} +\ell p_0 N^{(q)}$,
ensures the relativistic covariance of the
torsionful conservation law $(k \oplus p) \oplus q = 0$.
For what concerns the $0$-th component one easily finds
that
\begin{eqnarray}
&& \!\!\!\!\!\!\!\!\!\!\!\!\!\!\!\!\!\!\!\!\!\!\!\!\!\!\!\!\!\!\!\!\!
[N^{(k)}+N^{(p)}+N^{(q)} + \ell k_0 N^{(p)} + \ell k_0 N^{(q)}
+\ell p_0 N^{(q)}, ((k \oplus p) \oplus q)_0]
 =
 \nonumber\\
 && =  [N^{(k)}+N^{(p)}+N^{(q)} + \ell k_0 N^{(p)} + \ell k_0 N^{(q)}
+\ell p_0 N^{(q)}, k_0 + p_0 + q_0 ]=\nonumber\\
 \!\!\!\!\!\!\! \!\!\!\!\!\!\! \!\!\!\!\!\!\!
 & & = k_1+ p_1 +q_1+
 \ell k_0 p_1 + \ell k_0 q_1 + \ell p_0 q_1 =0
\label{covatritimetorsy}
\end{eqnarray}
 And similarly for the $1$-component one finds
that
\begin{eqnarray}
&& \!\!\!\!\!\!\!\!\!
[N^{(k)}+N^{(p)}+N^{(q)} + \ell k_0 N^{(p)} + \ell k_0 N^{(q)}
+\ell p_0 N^{(q)}, ((k \oplus p) \oplus q)_1]
 = \nonumber\\
 && =  [N^{(k)}+N^{(p)}+N^{(q)} + \ell k_0 N^{(p)} + \ell k_0 N^{(q)}
+\ell p_0 N^{(q)},
k_1 + p_1 + q_1 + \ell k_0 p_1
+ \ell k_0  q_1 + \ell p_0 q_1 ]=\nonumber\\
 & &
=  k_0+\ell k_0^2 + \frac{\ell}{2} k_1^2+ p_0+\ell p_0^2  + \frac{\ell}{2} p_1^2  +
 q_0+\ell q_0^2 + \frac{\ell}{2} q_1^2
+ 2 \ell k_0 p_0 + 2 \ell k_0 q_0 + 2 \ell p_0 q_0
+  \ell k_1 p_1 +  \ell k_1 q_1
+\ell p_1 q_1 = \nonumber\\
&&= k_0+p_0 +q_0 +\ell  (p_0 + k_0 + q_0)^2 + \frac{\ell}{2} (p_1 + k_1 +q_1)^2 = 0
~~~~~~~~~~~~~~~~~~~~~~~~~~~~~~~~~~~~~~~~~~~~~~~~~~~~~
\label{covatrispacetorsy}
\end{eqnarray}

So I have exposed a description of boosts that provides a satisfactory
notion of deformed Lorentz symmetry on the $\kappa$-momentum space.
And this again should be listed among the successes of the ``golden rule",
since, as shown above, the $\kappa$-momentum space does fit the demands of
the golden rule.

\subsection{Deformed Lorentz symmetry on a momentum space with torsion
and unmodified dispersion}\label{kappadue}
As a second torsionful example of doubly-special-relativity
type
deformation of Lorentz symmetry on momentum space, suitable also for
testing the ``golden rule", I take another variant of the$\kappa$-momentum space.
As mentioned the  $\kappa$-momentum space of the previous subsection
is inspired by studies of the $\kappa$-Poincar\'e Hopf algebra,
and specifically the work of Majid and Ruegg on the so-called ``Majid-Ruegg
basis of $\kappa$-Poincar\'e", and indeed we ended up
with adopting in the previous subsection the MR connection/composition law.
The second version of $\kappa$-momentum space which I now intend to
consider is inspired by a different choice of generators for
the $\kappa$-Poincar\'e Hopf algebra,
a choice of generators proposed in Ref.~\cite{gacAlessandraFrancesco}
which is obtained from the Majid-Ruegg generators by acting with
a ``change of basis"~\cite{kpoinap,gacAlessandraFrancesco}.

The Majid-Ruegg basis has been most frequently adopted in the $\kappa$-Poincar\'e
literature, mostly because some of the Hopf-algebraic manipulations
one is interested in doing turn out to be
particularly simple when using the Majid-Ruegg basis.
All this however is of little interest for my purposes here, and instead
I shall notice that the basis proposed by Agostini, D'Andrea and myself
in Ref.~\cite{gacAlessandraFrancesco}
provides inspiration for a description of the geometry
of momentum space with several intriguing features.
Reasoning just as done in Ref.~\cite{anatomy} for deriving $\kappa$-momentum space
from the Majid-Ruegg basis one can easily obtain from the basis
proposed in Ref.~\cite{gacAlessandraFrancesco}
a formulation of $\kappa$-momentum space which (in leading order) is ultimately
characterized by the undeformed
on-shell relation
\begin{equation}
m^2 = p_0^2 - p_1^2
\label{metrictorsyBBB}
\end{equation}
and a torsionful momentum-space affine connection
such that
\begin{equation}
(k \oplus p)_1 = k_1 + p_1 + \frac{\ell}{2} k_0 p_1 - \frac{\ell}{2} p_0 k_1  ~,~~~
(k \oplus p)_0 = k_0 + p_0~,
\label{connectiontorsyBBB}
\end{equation}
which in the following I shall label as the ``AAD composition law"
as a quick pointer to Ref.~\cite{gacAlessandraFrancesco}.

It is interesting to note that\footnote{Another noteworthy property
of (\ref{connectiontorsyBBB}) is $(p \oplus p)_\mu = 2 p_\mu$ (trivial composition
of ``parallel momenta").}
the antipode that follows from this AAD composition law
is trivial
\begin{equation}
(\ominus p)_\mu = - p_\mu ~,
\label{antipodetorsyBBB}
\end{equation}
which, especially when combined with the fact that the on-shell relation
is undeformed, renders this formalization of a $\kappa$-momentum space
particularly simple to handle, while preserving all the elements
of complexity of a torsionful momentum space.
The simple structure of the AAD composition law is also manifest
to some extent when several different momenta are combined,
as in the case of the composition of 3 momenta, which is given by
\begin{equation}
[(k \oplus p) \oplus q]_1 = k_1 + p_1 + q_1
+ \frac{\ell}{2} k_0 p_1 - \frac{\ell}{2} p_0 k_1
+ \frac{\ell}{2} k_0 q_1 - \frac{\ell}{2} q_0 k_1
+ \frac{\ell}{2} p_0 q_1 - \frac{\ell}{2} q_0 p_1  ~,~~~
[(k \oplus p) \oplus q]_0 = k_0 + p_0 + q_0
\label{tribodytorsyBBB}
\end{equation}

Before discussing the deformed-Lorentz-symmetry relativistic issues,
let me again pause for looking at this scenario with undeformed on-shell relation
and AAD composition law from the viewpoint of the ``golden rule"
derived in Sec.~\ref{goldenrulesec}.
It is easy to verify that also the combination of undeformed
on-shell relation
and AAD composition law
fits the demands of the ``golden rule".
In fact,
using again the parametrization introduced
in Sec.~\ref{goldenrulesec},
one sees that
undeformed
on-shell relation
and AAD composition law give $\alpha_1 = 0$, $\alpha_2 = 0$,
$\beta_1 = 0$, $\beta_2 = 0$,
$\gamma_1 = \ell/2$, $\gamma_2 = -\ell/2$, and therefore
indeed the golden rule is satisfied: $\alpha_1
+ \alpha_2 + \beta_1 + \beta_2
- \gamma_{{\!\!\!~}_1} - \gamma_{{\!\!\!~}_2} = 0$.\\
Once again, if we trust the ``golden rule" as (not only a necessary but also) a sufficient
condition for DSR-compatibility we must then expect
that this scenario with undeformed  on-shell relation
and AAD composition law should admit
 a satisfactory notion
of deformed Lorentz symmetry. And indeed this is what I shall now
show.

As explained in the previous subsection, even for such torsionful momentum
spaces the  description of doubly-special-relativity-type
deformed boosts  on momenta
of a single particle is not challenging. For this case with undeformed
on-shell relation
and AAD composition law I propose
\begin{equation}
[N, p_0] =  p_1 + \frac{\ell}{2} p_0 p_1 ~,~~~
[N, p_1] =  p_0 + \frac{\ell}{2} p_0^2~,
\label{booststorsyBBB}
\end{equation}
which evidently ensures the
invariance of the on-shell relation:
\begin{equation}
[N, p_0^2 - p_1^2] =
2 p_0 (p_1 + \frac{\ell}{2} p_0 p_1) - 2 p_1 (p_0 + \frac{\ell}{2} p_0^2) = 0
\label{invariantshelltorsyBBB}
\end{equation}
Next I must deal again with the fact that
torsion inevitably affects how boosts act on
momenta obtained composing two or more single-particle momenta.\\
It is noteworthy that I find  that the same
prescription used for the other $\kappa$-Poincar\'e-inspired
torsionful case considered
in the previous subsection,
$$N^{(k \oplus p)} =N^{(k)}+N^{(p)} + \ell k_0 N^{(p)}~,$$
also works for the
the case I am considering in this subsection.

Let me start verifying this on $k \oplus p = 0$. For what concerns the $0$-th component
one easily finds
\begin{eqnarray}
[N^{(k)}+N^{(p)} + \ell k_0 N^{(p)}, (k \oplus p)_0]=
[N^{(k)}+N^{(p)} + \ell k_0 N^{(p)}, k_0 + p_0] &=& k_1
+\frac{\ell}{2} k_0 k_1
+p_1
+\frac{\ell}{2} p_0 p_1
+\ell k_0 p_1 \label{covatwotimetorsyBBB}\\
&=& k_1 +p_1
+\frac{\ell}{2} k_0 p_1
-\frac{\ell}{2} p_0 k_1
+\frac{\ell}{2} (k_0 + p_0) (k_1 + p_1)
= 0 ~,
\nonumber
\end{eqnarray}
where on the right-hand side I of course
used the conservation law itself.\\
Similarly for the $1$-component of $k \oplus p = 0$
it turns out that
\begin{eqnarray}
[N^{(k)}+N^{(p)} + \ell k_0 N^{(p)}, (k \oplus p)_1]
 &=& [N^{(k)}+N^{(p)} + \ell k_0 N^{(p)}, k_1 + p_1
 + \frac{\ell}{2} k_0 p_1 - \frac{\ell}{2} p_0 k_1 ]=\nonumber\\
 &= & k_0+ \frac{\ell}{2} k_0^2
 + p_0+ \frac{\ell}{2} p_0^2
 + \ell k_0 p_0  \nonumber\\
 &= & k_0+p_0 + \frac{\ell}{2} (k_0 + p_0)^2 = 0
~~~~~~~~
\label{covatwospacetorsyBBB}
\end{eqnarray}
where again on the right-hand side I
used the conservation law $(k \oplus p)_\mu = 0$ itself
(and took again into account that
I am working at leading order in $\ell$).

Consistently with the style of analysis I adopted throughout this manuscript
let me double-check this consistency between my proposal for boosts
and the choice of undeformed on-shell relation
and AAD composition law by also considering the case of a 3-particle conservation
law, of the form $(k \oplus p) \oplus q = 0$.
For what concerns the $0$-th component one easily finds
that
\begin{eqnarray}
&& \!\!\!\!\!\!\!\!\!\!\!\!\!\!\!\!\!\!\!\!\!\!\!\!\!\!\!\!\!\!\!\!\!
[N^{(k)}+N^{(p)}+N^{(q)} + \ell k_0 N^{(p)} + \ell k_0 N^{(q)}
+\ell p_0 N^{(q)}, ((k \oplus p) \oplus q)_0]
 =
 \nonumber\\
 && =  [N^{(k)}+N^{(p)}+N^{(q)} + \ell k_0 N^{(p)} + \ell k_0 N^{(q)}
+\ell p_0 N^{(q)}, k_0 + p_0 + q_0 ]=\nonumber\\
 \!\!\!\!\!\!\! \!\!\!\!\!\!\! \!\!\!\!\!\!\!
 & & =
 k_1
+\frac{\ell}{2} k_0 k_1
+p_1
+\frac{\ell}{2} p_0 p_1
+q_1
+\frac{\ell}{2} q_0 q_1
+\ell k_0 p_1+\ell k_0 q_1+\ell p_0 q_1 \nonumber\\
&=& k_1 +p_1 +q_1
+\frac{\ell}{2} k_0 p_1
-\frac{\ell}{2} p_0 k_1
+\frac{\ell}{2} k_0 q_1
-\frac{\ell}{2} q_0 k_1
+\frac{\ell}{2} p_0 q_1
-\frac{\ell}{2} q_0 p_1
+\frac{\ell}{2} (k_0 + p_0 + q_0) (k_1 + p_1 + q_1)
= 0 ~.
\label{covatritimetorsyBBB}
\end{eqnarray}
 And similarly for the $1$-component one finds
that
\begin{eqnarray}
&& \!\!\!\!\!\!\!\!\!
[N^{(k)}+N^{(p)}+N^{(q)} + \ell k_0 N^{(p)} + \ell k_0 N^{(q)}
+\ell p_0 N^{(q)}, ((k \oplus p) \oplus q)_1]
 = \nonumber\\
 && =  [N^{(k)}+N^{(p)}+N^{(q)} + \ell k_0 N^{(p)} + \ell k_0 N^{(q)}
+\ell p_0 N^{(q)},
k_1 + p_1 + q_1
+ \frac{\ell}{2} k_0 p_1 - \frac{\ell}{2} p_0 k_1
+ \frac{\ell}{2} k_0 q_1 - \frac{\ell}{2} q_0 k_1
+ \frac{\ell}{2} p_0 q_1 - \frac{\ell}{2} q_0 p_1]=\nonumber\\
 & &
=  k_0+ \frac{\ell}{2} k_0^2+ p_0+ \frac{\ell}{2} p_0^2  +
 q_0+ \frac{\ell}{2} q_0^2
+  \ell k_0 p_0 +  \ell k_0 q_0 +  \ell p_0 q_0
 = \nonumber\\
&&= k_0+p_0 +q_0 +\frac{\ell}{2} (p_0 + k_0 + q_0)^2
 = 0
~~~~~~~~~~~~~~~~~~~~~~~~~~~~~~~~~~~~~~~~~~~~~~~~~~~~~
\label{covatrispacetorsyBBB}
\end{eqnarray}

So once again the golden rule was successfully tested: for the choice of
 undeformed on-shell relation
and AAD composition law the golden rule is satisfied and indeed I managed to expose
a description of doubly-special-relativity-type deformed boosts that ensures
the invariance of the undeformed on-shell relation and the covariance
of the conservation laws
obtained from the AAD composition law.

\section{On the momentum-space metric}\label{newmetric}
I have here set the stage for analyzing whether a given geometry of momentum space,
as codified in an on-shell (dispersion) relation and in a law of conservation
of momenta at interactions, may or may not be DSR-compatible, {\it i.e.}
compatible with a relativistic
formulation of theories on that momentum space.\\
I found that the availability of a relativistic picture requires
enforcing the ``golden rule", {\it i.e.} a certain
consistency between the on-shell relation and the conservation laws
({\it e.g.} between the parameters $\alpha_1,\alpha_2$
and the parameters $\beta_1,\beta_2,\gamma_1,\gamma_2$).\\
An attractive result of such an analysis would have been to find that
this relativistic consistency of combined choices of
on-shell relation and conservation laws
could have a simple geometric meaning: for example, a case such that given
a certain metric on momentum space (used to codify the on-shell relation)
the type of affine connections on momentum space (used to codify the
conservation laws) that would allow a relativistic picture would have been,
say, metric connections or perhaps non-metric connections of a certain specific type.
However, I have not managed to find a way
to use the notion of metric and affine connection on momentum space
introduced in Refs.~\cite{prl,grf2nd} for providing such a simple characterization.

There may well be such a characterization which I still did not notice,
and in any case the availability of such a characterization can at best
be viewed as desirable, but not necessarily available.
 It is interesting however to speculate
about possible alternative geometric pictures, other ways to associate
a metric to the on-shell relation and/or to associate an affine connection
to the conservation laws. A suitable alternative geometric picture
might even lead to a very intuitive geometric characterization of what it takes
for a momentum space to be DSR-compatible.\\
I do not have any particularly compelling proposal at present, but
let me illustrate the type of alternative ``geometric picture of momentum
space" that I have in mind by motivating a possibility
in which the affine
connection is still associated with the conservation laws exactly as
prescribed in Refs.~\cite{prl,grf2nd}, but one would link the metric to the
on-shell relation in a way that differs from the one used in Refs.~\cite{prl,grf2nd}.
This alternative proposal finds part of its inspiration in a field-theory
argument which I am now going to propose.

Let me start by reminding the reader about the known fact that classical
field theories in Minkowski spacetime can be easily formulated
as classical field theories on a momentum space with Minkowski metric.
For example, a free massless scalar field
is described by the momentum-space action
$$ -  \int d^4 k \int d^4 p~ {\widetilde \phi}(k)  \eta^{\mu \nu} k_\mu  p_\nu
  ~ {\widetilde \phi}(p) \delta(k+p)$$
  where $\eta^{\mu \nu}$ is to be viewed here as a Minkowski metric
  on momentum space, and from the viewpoint of Refs.~\cite{prl,grf2nd} 
  the ordinary linear sum of momenta, $k+p$,
  codifies the trivial Levi-Civita connection of the Minkowski metric.\\
Taking as  ``spacetime field" the $\phi(x)$ obtained from the momentum-space
field ${\widetilde \phi}(p)$ by anti-Fourier transform,
${\widetilde \phi}(p) = \int d^4 x
 \phi(x) e^{-ikx}$, one indeed obtains with few steps of simple derivation
 the standard spacetime description of a free massless scalar field:
$$\int  d^4x ~\int  d^4y ~~ \eta^{\mu \nu} ~\left(\partial_\mu \phi(x)\right)
 ~ \left( \partial_\nu \phi(y) \right)~ \delta^4(x-y)~.$$

In light of this it appears natural to assume that
on a momentum space which does not have Minkowski geometry
 a free massless scalar field should be described in terms of the Lagrangian density
\begin{equation}
 -  {\widetilde \phi}(k) ~ g^{\mu \nu}(p) \, k_\mu  p_\nu
 ~ {\widetilde \phi}(p) \delta(k \oplus p)
 \label{preDalambert}
 \end{equation}
 where $g^{\mu \nu}$ is the momentum-space metric and $\oplus$ is
 the composition law obtained from the momentum-space affine connection
 following the corresponding prescription of Refs.~\cite{prl,grf2nd}.\\
For my purposes it is valuable to notice that
in particular this should allow one to integrate over $k$ finding
$$ 
 -  {\widetilde \phi}(\ominus p)
  ~ g^{\mu \nu}(p) \,  (\ominus p)_\mu   p_\nu
 ~ {\widetilde \phi}(p)~,$$
where $\ominus$ denotes again the antipode of the composition law.
And evidently this is suggesting that the on-shell relation
(for massless particles) could be viewed as involving the metric on
momentum space together with a momentum and its antipode:
\begin{equation}
0 = - g^{\mu \nu}(p) \, (\ominus p)_\mu p_\nu
\label{newmaster}
\end{equation}
So following this particular line of analysis (but several other possibilities
should perhaps be considered) I was led to contemplating a role for
the momentum-space metric in the form of the on-shell relation
which is alternative to the one of Ref.~\cite{prl}.

At present I do not see any particularly compelling argument
to prefer one or another way to link momentum-space metric and on-shell relation.
One little observation which could be viewed as favoring the alternative (\ref{newmaster})
can be based on the two examples of ``$\kappa$-Poincar\'e-inspired
momentum spaces" which I considered in Subsections  \ref{kappauno} and \ref{kappadue}.
To see this let us start with the ``MR $\kappa$-momentum space" of
Subsection  \ref{kappauno}: there the on-shell relation is
\begin{equation}
m^2 = p_0^2 - p_j^2 + \ell p_0 p_j^2
\label{onshellkappauno}
\end{equation}
and the antipode is given by
\begin{equation}
(\ominus p)_j = - p_j + \ell p_0 p_j  ~,~~~
(\ominus p)_0 = - p_0~,
\label{antipodekappauno}
\end{equation}
and intriguingly one can obtain the on-shell relation
from the antipode
by using the Minkowski metric:
$$0 = - \eta^{\mu \nu} (\ominus p)_\mu p_\nu=
-(-p_0) p_0
+(-p_j + \ell p_0 p_j) p_j
=
p_0^2 - p_j^2 + \ell p_0 p_j^2$$
And the same observation also applies to
the  ``AAD $\kappa$-momentum space" of Subsection \ref{kappadue},
since there the (leading-order)
 on-shell relation is undeformed
$$m^2 = p_0^2 - p_j^2$$
and also the antipode
is trivial (in spite of the nontriviality of the AAD composition
law (\ref{connectiontorsyBBB})),
$$(\ominus p)_\mu = - p_\mu~,$$
so that evidently one obtains the undeformed on-shell relation
from a trivial antipode using the prescription (\ref{newmaster})
with a Minkowski metric on momentum space:
$$- \eta^{\mu \nu} (\ominus p)_\mu p_\nu=
-(-p_0) p_0
+(-p_j ) p_j
=
p_0^2 - p_j^2~.$$

It is still not clear to me  whether this observation should tell us something
about the proper notion of how to connect the on-shell relation with
momentum-space metric and/or tell us something that
specifically holds for
the $\kappa$-Poincar\'e framework. It may even turn out to be
ultimately meaningless, but it appeared to be sufficiently intriguing
to be worth bringing it
to the attention of readers of this manuscript.

\section{A theorem without formulas}\label{theoremsec}
As I am getting close to conclude this study, let me pause for a brief aside, just to show
that the criterion I introduced in Sec.~II, and the associated ``golden rule",
must be understood from a broader perspective as
a particular application, of significant practical value,
of a simple relativistic theorem which I shall here
state and provide the elementary (no formulas) proof for.

The compact (uncareful) version of the theorem states that {\it in a relativistic
theory one cannot have one-particle energy thresholds characterizing whether or not
a given process is kinematically allowed}.

\medskip

\noindent
The {\underline{hypotheses}} of the theorem are:

\noindent
$\bullet$ The theory is relativistic, in the sense that there is no preferred-frame description,
and reproduces special relativity in the low-energy limit (limit in which all particles
have small energies).

\noindent
$\bullet$ The threshold energy of interest concerns a single specific particle
 taking part in a process (say, the muon in the out state of $\pi^+ \rightarrow \mu^+ \nu_\mu$)
 and marks the separation between a kinematical regime where the process is allowed
 and a kinematical regime where the process is not allowed
 (say, when the muon has $E_\mu > E_{threshold}$ the process $\pi^+ \rightarrow \mu^+ \nu_\mu$
 is/is not allowed, whereas for  $E_\mu < E_{threshold}$ the process $\pi^+ \rightarrow \mu^+ \nu_\mu$
 is not/is allowed).

\noindent
$\bullet$ The theory may or may not have an observer-independent energy scale,
but if it does the theorem anyway focuses on considering cases where
the value of the threshold energy of interest is not an invariant under the
boost transformations of the relativistic theory.

\noindent
$\bullet$ Particle reactions are objective physical processes.

\medskip

\noindent
A slightly more careful  {\underline{statement of the theorem}} is:

\noindent
$\diamond$ It is not possible to have threshold-energy laws of the type described
in these hypotheses,
in a relativistic theory
fulfilling these hypotheses.

\medskip

\noindent
{\underline{PROOF:}} I can just show that such a threshold-energy law $E_X > E_{threshold}$
for a particle of type $X$, involved in a process of type $X+A_1+\dots +A_n \rightarrow B_1 + \dots + B_m$
or of type $A_1+\dots +A_n \rightarrow X+ B_1 + \dots + B_m$,
is not a relativistic law, within the hypotheses of the theorem.
Consider then observer Alice according to which $E_X$ is just above
threshold, $E_X^{(Alice)} = E_{threshold} + \epsilon$, so that according to Alice the process
is kinematically
allowed (respectively kinematically not allowed). Since the hypotheses ensure that $E_X$ is not
invariant under boosts (even in cases where there is some energy scale which is invariant under boosts)
we can safely assume that under the action of an appropriate boost transformation
we can reach from Alice an observer Bob for whom the energy of that same particle
of type $X$ is below threshold: $E_X^{(Bob)} = E_{threshold} - \epsilon$, and for whom then,
since the hypotheses insist on the case of a relativistic theory, that same
process is instead kinematically not
allowed (respectively kinematically allowed). All this evidently comes into contradiction
with the objectivity of particle reactions. {\underline{End of proof.}}

\medskip
The careful reader can easily recognize how this applies in particular to the
case of a threshold law for the energy of the photon in the process $\gamma \rightarrow e^+ e^-$,
and the associated requirement that one should enforce a ``no-photon-decay-switch-on constraint".\\
And it is only slightly less immediately obvious that also
my  ``no-pair-production-switch-off constraint" for $\gamma + \Gamma \rightarrow e^+ e^-$
is a particular case of this theorem:
if there was a lower threshold for the energy of the soft photon $\gamma$ below which
the process $\gamma + \Gamma \rightarrow e^+ e^-$ was no longer allowed kinematically
(because hard photons $\Gamma$ of no matter how high energy could still not produce
electron-positron pairs in interactions with such soft photons $\gamma$) we would
indeed be coming in contradiction with the theorem.

\medskip
Stating this simple, nearly self-evident, theorem may appear to be a bit of a
redundant overkill
in light of the discussion here reported, but it was evidently not in the minds
of the authors (a few) who have sought ``anomalous particle-decay thresholds" 
in DSR-compatible pictures,
since in all such contexts the discussion I gave here of the ``no-photon-decay-switch-on constraint"
for $\gamma \rightarrow e^+ e^-$ would evidently also apply up to trivial generalization.
And it was evidently also not in the minds
of the (several) authors who have sought a DSR-compatible 
photopion-production-switch-off threshold
as a way to ``explain" the once presumed
cosmic-ray GZK-threshold anomaly (for which now there is no longer any evidence~\cite{cosmicrayUPDATE}):
such a photopion-production-switch-off threshold is necessarily obstructed in DSR-compatible
frameworks for the same reasons which here led me to exclude a
pair-production-switch-off threshold.

I myself had put forward the half of the theorem which is essentially codified by
the ``no-photon-decay-switch-on constraint" already in Ref.~\cite{gacnewjourn}
(also see the later study in Ref.~\cite{sethmajor}, and the very recent related
study in Ref.~\cite{operaDSR},
essentially generalizing from photons to neutrinos the no-decay-switch-on constraint),
but I put in focus only rather recently the other half of the theorem, the one
which is essentially contained in my new ``no-pair-production-switch-off constraint".

While the theorem renders obvious the fact
that satisfying these two constraints is a {\underline{necessary}} condition for DSR-compatibility,
it remains of course uncertain whether satisfying these two constraints
(enforcing the ``golden rule",
within a chosen parametrization) can really provide a {\underline{sufficient}} condition for
DSR-compatibility, as the few tests I here reported appear to suggest.

\section{Closing remarks}\label{closingsec}
The recently-proposed ``relative-locality framework" has the potential of
producing several new physical pictures and of accommodating,
reformulating them accordingly,
some previously existing proposals.
I have here show that, as for other areas of quantum-gravity-inspired research,
the DSR concept and the techniques of analysis that have been developed
in order to explore it, prove valuable also for the analysis of
this powerful framework.\\
I have here established
that some relative-locality momentum spaces necessarily require a formulation
of theories with a preferred frame, and I provided a robust and technically simple
strategy for identifying such momentum-space geometries.
The fact that, within a chosen parametrization of momentum space geometry,
my criteria produce a simple algorithmic classification of momentum-space
geometries that require a preferred-frame description
is particularly convenient for future applications of the results here reported.\\
The conjecture that my criterion (and the associated ``golden rule")
might be not only necessary but also sufficient for DSR-compatibility
found some support in the results here reported in
Sections \ref{dsr1sec} and \ref{torsionsec},
but evidently needs to be investigated in greater generality.\\
For what concerns specific models, especially
the results I here obtained on the DSR-compatibility
of ``$\kappa$-momentum spaces"
may be of particular interest, since these are evidently among the most
natural ``laboratories" for a first phase of exploration
of the implications of the relative-locality framework.

I have here studied DSR-compatibility of theories
on a relative-locality momentum space, but focusing exclusively
on momentum-space properties.
This is the most fundamental level at which to
investigate the availability of a deformed-Lorentz-symmetry picture,
since in the ``relative-locality framework"
momentum space is primitive and spacetime is only a derived entity~\cite{prl,grf2nd}.
But of course conceptually the most intriguing questions are to be expected
on the side
of the actual relativity of spacetime locality, and it will be extremely
interesting to investigate the interplay between (deformed-)boost invariance
and the relativity of spacetime locality.
I have here postponed this task, also assuming that my effort of rendering
more robust our understanding of boost transformations on a momentum space
with nontrivial geometry would provide solid ground for those further studies.

As I was in the final stages of preparation of this manuscript
I became aware of the papers in Refs.~\cite{flagiuKAPPAPRL,flajosePRL},
parts of which were devoted to issues pertaining (deformed) Lorentz
transformations with relative locality.
None of the points I here made about establishing the general criteria
for consistency of the momentum-space geometry with
(possibly deformed) Lorentz invariance is found in Refs.~\cite{flagiuKAPPAPRL,flajosePRL}.
Instead Refs.~\cite{flagiuKAPPAPRL,flajosePRL} each
take a specific scenario for momentum space and a specific scenario
for boost transformations, attempting to also establish a few starting
points for the analysis of the implications for relative spacetime locality.
Perhaps the aspect of these two manuscripts which could be most valuable
for the future development of the results I here reported
is the observation in Ref.~\cite{flagiuKAPPAPRL}
suggesting that for a proper description of finite (deformed-)boost transformations
in a relative-locality framework it might be
necessary to introduce a
dependence of the rapidity parameter on the particles momenta. I have here confined
my analysis at the complementary level of symmetry generators, setting the stage
 for describing the case of infinitesimal boost transformations. At least in
 the $\kappa$-momentum space I here considered in Subsection~\ref{kappauno}, which
 was the momentum space considered in Ref.~\cite{flagiuKAPPAPRL}, it might
 be fruitful to contrast and perhaps to complement my observations on symmetry
 generators and the observations reported in Ref.~\cite{flagiuKAPPAPRL}
 for the rapidity parameters.

Another issue for future studies, perhaps the most significant one,
concerns the interplay between deformed Lorentz symmetry and
 the implications of relative spacetime locality for cases with
 several causally-connected events. One should achieve a consistent relativistic description
 applicable to the case of pairs of distant and relatively boosted
 observers. As shown in Ref.~\cite{anatomy} even just studying distant observers in
 relative rest
 is already rather challenging when relative spacetime locality is taken into account for
 causally-connected events, so the generalization to
distant and relatively boosted
 observers may prove very challenging.

\bigskip
\bigskip

\noindent
{\large{\bf Acknowledgements}}

\medskip

I am grateful to my fellow explorers of the relative-locality momentum space,
Laurent Freidel, Jerzy Kowalski-Glikman and Lee Smolin, for the enjoyment of so many
wonderful discussions, some of which helped this work considerably. I am similarly grateful
to my fellow members of "team what about Bob", Niccolo' Loret,
Marco Matassa, Flavio Mercati
and Giacomo Rosati, who also contributed a great deal to my views
on the subject here discussed.

\end{document}